\documentclass{IEEEtran}
\usepackage{cite}
\usepackage{amsmath,amssymb,amsfonts}
\usepackage{algorithmic}
\usepackage{graphicx}
\usepackage{textcomp}
\usepackage{longtable}

\usepackage{makecell}
\usepackage{array}
\usepackage{hyperref}

\usepackage{tikz}
\usepackage{forest}
\usetikzlibrary{trees,positioning,shapes,shadings,shadows,arrows.meta,calc,fit}
\usepackage{eqparbox}

\usepackage{booktabs}
\usepackage{pifont}
\usepackage{rotating}

\newcommand*\rot{\rotatebox{90}}

\newcommand*\OK{\ding{51}}
\newcommand*\NOK{\ding{55}}

\newcolumntype{P}[2]{%
  >{\begin{turn}{#1}\begin{minipage}{#2}\small\raggedright\hspace{0pt}}l%
  <{\end{minipage}\end{turn}}%
}

\usepackage{fourier}
\usepackage{colortbl}
\usepackage{cleveref}

\usepackage[caption=false, font=footnotesize]{subfig}

\let\citet\cite

\begin{document}

\title{Audio-Language Datasets of Scenes and Events: \\ A Survey}
\author{%
\IEEEauthorblockN{%
Gijs Wijngaard\IEEEauthorrefmark{1}, Elia Formisano\IEEEauthorrefmark{2}, Michele Esposito\IEEEauthorrefmark{2} and Michel Dumontier\IEEEauthorrefmark{1}%
}

\IEEEauthorblockA{\IEEEauthorrefmark{1} 
Department of Advanced Computing Sciences,
Faculty of Science and Engineering,
Maastricht University \\
}%
\IEEEauthorblockA{\IEEEauthorrefmark{2} 
Department of Cognitive Neuroscience,
Faculty of Psychology and Neuroscience,
Maastricht University
}%
}

\maketitle

\markboth{Wijngaard \MakeLowercase{\textit{et al.}}: Audio-Language Datasets of Scenes and Events: A Survey}
{Wijngaard \MakeLowercase{\textit{et al.}}: Audio-Language Datasets of Scenes and Events: A Survey}

\begin{abstract}
Audio-language models (ALMs) generate linguistic descriptions of sound-producing events and scenes. Advances in dataset creation and computational power have led to significant progress in this domain. This paper surveys 69 datasets used to train ALMs, covering research up to September 2024 (\href{https://github.com/GLJS/audio-datasets}{github.com/GLJS/audio-datasets}). It provides a comprehensive analysis of datasets origins, audio and linguistic characteristics, and use cases. Key sources include YouTube-based datasets like AudioSet with over two million samples, and community platforms like Freesound with over 1 million samples. Through principal component analysis of audio and text embeddings, the survey evaluates the acoustic and linguistic variability across datasets. It also analyzes data leakage through CLAP embeddings, and examines sound category distributions to identify imbalances.  Finally, the survey identifies key challenges in developing large, diverse datasets to enhance ALM performance, including dataset overlap, biases, accessibility barriers, and the predominance of English-language content, while highlighting opportunities for improvement.
\end{abstract}


\section{Introduction}
Audio-Language learning, also referred to as Audio-Text learning, is a rapidly growing research field that focuses on processing, understanding and describing sounds using natural language. Audio-language models fall under the broader category of Multimodal Large Language Models, which can process and generate responses based on diverse input modalities \cite{yinSurveyMultimodalLarge2024}.  

Advancements in machine learning and the increasing availability of datasets that pair sounds with corresponding textual descriptions are driving the growth of audio-language learning. The field has witnessed a surge in activity due to the development of large language models (LLMs), which have inspired new research directions \cite{zhaoSurveyLargeLanguage2024, naveedComprehensiveOverviewLarge2024}. For example, natural language processing models based on the Transformer architecture \cite{vaswaniAttentionAllYou2017} have been adapted for audio-language Learning in various settings \cite{meiAutomatedAudioCaptioning2022, xuStatusQuoContemporary2024, nagSystematicLiteratureReview2021, wangCrossModalRetrievalSystematic2024, latifSparksLargeAudio2023a}. Additionally, techniques such as contrastive learning - designed to differentiate similar and dissimilar data pairs - and transfer learning have led to remarkable performance enhancement across various audio-language tasks \cite{deshmukhPengiAudioLanguage2023, elizaldeNaturalLanguageSupervision2024, wuLargeScaleContrastiveLanguageAudio2023, anupriyaTransferLearningAudio2023, paissanTinyCLAPDistillingConstrastive2024}.

The availability of high-quality datasets has been another key driver of progress in this field. Two pivotal milestones were the release of the Freesound platform in 2013 \cite{fontcorberaFreesoundTechnicalDemo2013} and AudioSet in 2017 \cite{gemmekeAudioSetOntology2017}. These resources have enabled training models for a wide range of sounds and tasks. Since then, numerous studies have followed \cite{kimAudioCapsGeneratingCaptions2019, drossosClothoAudioCaptioning2020, wuLargeScaleContrastiveLanguageAudio2023, meiWavCapsChatGPTAssistedWeaklyLabelled2023}, leading to the development and refinement of datasets tailored to the needs of specific audio-language models \cite{gongListenThinkUnderstand2023, deshmukhPengiAudioLanguage2023, chuQwenAudioAdvancingUniversal2023}. 

Recent studies indicate scaling laws for the development of foundation models \cite{kaplanScalingLawsNeural2020, jordanhoffmannTrainingComputeOptimalLarge2022}: a correlation between model performance and the amount of data and training time these models use. In Audio-Language Learning, for example, the ESC-50 audio classification dataset requires around 2 million audio-text data points for a zero-shot model to match human parity \cite{yanpengzhaoConnectingDotsAudio2022}. An issue in retrieving audio-language data is that the amount of training audio data available is inevitably less than that for natural language processing tasks \cite{takeuchiEffectsWordfrequencyBased2020}. Current audio datasets are mostly based on sound effect databases, YouTube, or Freesound for their audio, whereas LLMs are trained on a large portion of the open Internet \cite{raffelExploringLimitsTransfer2023}. A limited training data set has presented the issue of \textit{distributional shift}: a mismatch between training data and production data, which makes a model poorly equipped for real-world data \cite{changContextPromptEditing2023}. Models trained with contrastive learning (CLAP) \cite{wuLargeScaleContrastiveLanguageAudio2023, elizaldeNaturalLanguageSupervision2024}, benefit from large and diverse datasets for better performance in audio-language domains \cite{kimOvercomingDataShortage2023, liuCL4ACContrastiveLoss2021a, wuAudioTextModelsNot2023, noriyCLARAMultilingualContrastive2023, elizaldeNaturalLanguageSupervision2024}.

A recent development in audio-language datasets is the use of LLMs for annotating, generating or augmenting training data for audio-language models (e.g. \cite{wuLargeScaleContrastiveLanguageAudio2023, meiWavCapsChatGPTAssistedWeaklyLabelled2023, gongListenThinkUnderstand2023, primusCpjkusSubmissionTask2023, kongImprovingTextAudioModels2024, kreukAudioGenTextuallyGuided2023, robinsonTransferableModelsBioacoustics2024, xiaoEnsembleSystemsContrastive2023, primusAdvancingNaturalLanguageBased2023,kimOvercomingDataShortage2023, sunAutoACDLargescaleDataset2024, ghoshRecapRetrievalaugmentedAudio2024, guLanguagebasedAudioRetrieval2024, chenSjtuthuAutomatedAudio2024, kulikTakeItGranted2024}, see also \cite{zhouSurveyDataAugmentation2024}). 
LLM-generated sound captions offer a way to address the scarcity of high-quality, human-annotated sound datasets. Another complementary approach, as demonstrated in this work, involves utilizing a wider range of existing, diverse data sources.

This article surveys the available datasets in the audio-language domain, analyzing 69 datasets, significantly expanding upon previous surveys that reviewed 2 datasets \cite{wangCrossModalRetrievalSystematic2024}, 3 datasets \cite{meiAutomatedAudioCaptioning2022, xuStatusQuoContemporary2024, nagSystematicLiteratureReview2021}, 5 datasets \cite{latifSparksLargeAudio2023a} and 20 datasets \cite{triantafyllopoulosComputerAuditionTaskspecific2024}. The specific contributions of this work are twofold:
\begin{enumerate}
\item \textbf{Comprehensive Dataset Overview}: It provides a detailed overview of datasets used in the field of audio-language models, including relevant statistics (Table \ref{tab:datasets}) such as the average length of audio and text (captions), total hours, sources and application domains. 
\item \textbf{Data Leak Analysis:} It includes a data overlap analysis (Section \ref{sec:dupe-analysis}) assessing the extent of data leakage between datasets. This analysis is critical for combining datasets, developing new ones and properly evaluating zero-shot learning tasks. 
\end{enumerate}

"Zero-shot" learning, the capability of models to recognize sounds they were not explicitly trained on, is becoming a relevant aspect for evaluating audio-language models \cite{tangSALMONNGenericHearing2024, ghoshGAMALargeAudioLanguage2024a, huangMakeAudio2TemporalEnhanced2023, deshmukhPengiAudioLanguage2023}. However, achieving true zero-shot learning is challenging in practice. Pre-trained models may have already been exposed to similar data during training due to data leakage or overlap between datasets \cite{tavaresClassSeparabilityPitfalls2024}. For example, a model pre-trained on AudioSet \cite{gemmekeAudioSetOntology2017} may already have seen \textit{"dog barks"} sound samples, which raises the question of whether recognizing\textit{ "dog barks"} in a new dataset represents genuine zero-shot learning. This highlights the importance of thorough dataset analysis to understand the true capabilities and limitations of audio-language models and to ensure robust evaluation of zero-shot performance. The data leak analysis presented in this survey directly addresses these concerns by examining overlap between datasets.

Given the extensive scope of the audio-language field, this article focuses on two specific aspects. First, it examines the domain of sound and event machine listening, deliberately excluding speech and music modeling from its scope (for surveys on speech and music modeling, see \cite{mehrishReviewDeepLearning2023, prabhavalkarEndEndSpeechRecognition2024, latifSparksLargeAudio2023a, abhyankarSurveyMusicGenre2023}). Second, it distinguishes between different types of machine listening tasks. Traditional tasks, such as Sound Event Detection (SED) and Audio Tagging (AT) are limited to predicting within the scope of the labeled data on which they were trained. For example, a model trained on the AudioSet dataset can only predict labels from its 527 classes. In contrast, recent audio-language models utilize an encoder-decoder setup, integrating an audio understanding and representation model with a text understanding and representation model.
This survey focuses on such audio-language models and the tasks they support, including Automated Audio Captioning (AAC), Audio-Text Retrieval (ATR), Automated Audio Generation (AAG) and Text-to-Audio Retrieval (ATR). These tasks, along with others, are briefly introduced in Section \ref{sec:background}.

The survey is organized as follows: The Background section (Section \ref{sec:background}) introduces the field and its models.
It then reviews related work on audio-language research and describes the research methodology employed. A comprehensive overview of the identified datasets in the field is provided in Section \ref{sec:datasets}. The survey also includes an analysis of data leakage (Section \ref{sec:dupe-analysis}), followed by a discussion of bias and data quality in audio datasets (Section \ref{sec:discussion}). Finally, the paper concludes with a summary of findings and directions for future research (Section \ref{sec:conclusion}).

\begin{figure*}[t]
    \centering
    \caption{Audio-Language models: Audio-to-Language models (top row) process audio input to generate text output, while Text-to-Audio models (bottom row) process text input to manipulate or generate audio output.}
    \scalebox{0.75}{
    \begin{tikzpicture}[
        box/.style={draw, minimum width=2cm, minimum height=1.5cm, rounded corners},
        arrow/.style={->, >=latex, thick}
    ]
        \begin{scope}[xshift=0cm]
            \node (output1) at (0,4) {\small $\langle$sos$\rangle$ dog barking outside $\langle$eos$\rangle$};
            \node[box, fill=red!20] (dec1) at (0,2) {Text Decoder};
            \node[box, fill=gray!20] (latent1) at (0,0) {Latent Space};
            \node[box, fill=blue!20] (enc1) at (0,-2) {Audio Encoder};
            \node (input1) at (0,-4) {\includegraphics[width=1.8cm]{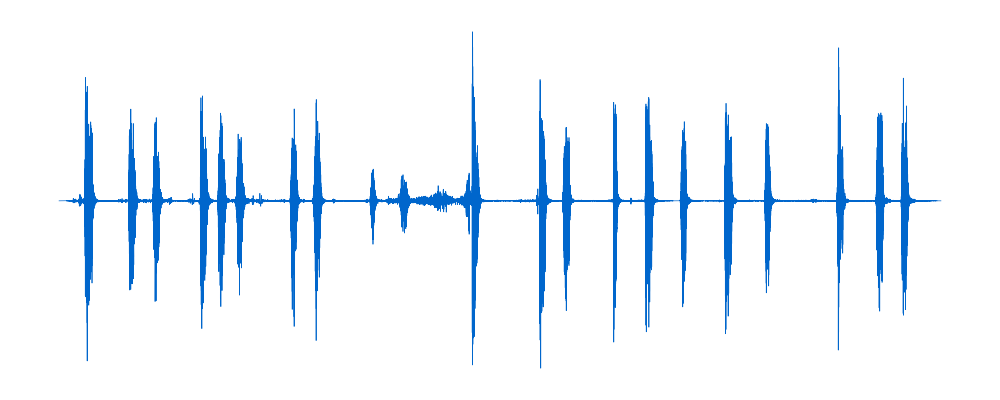}};
            
            \draw[arrow] (input1) -- (enc1);
            \draw[arrow] (enc1) -- (latent1);
            \draw[arrow] (latent1) -- (dec1);
            \draw[arrow] (dec1) -- (output1);
            
            \node[above, font=\sffamily] at (0,4.5) {Audio Captioning};
        \end{scope}

        \begin{scope}[xshift=6cm]
            \node (output2) at (0,4) {\small "Added bird chirping"};
            \node[box, fill=red!20] (dec2) at (0,2) {Difference Decoder};
            \node[box, fill=gray!20] (comp2) at (0,0) {Comparison Layer};
            \node[box, fill=blue!20] (enc2) at (-1.5,-2) {Audio Encoder 1};
            \node[box, fill=blue!20] (enc2b) at (1.5,-2) {Audio Encoder 2};
            \node (input2a) at (-1.5,-4) {\includegraphics[width=1.5cm]{wijng1}};
            \node (input2b) at (1.5,-4) {\includegraphics[width=1.5cm]{wijng1}};
            
            \draw[arrow] (input2a) -- (enc2);
            \draw[arrow] (input2b) -- (enc2b);
            \draw[arrow] (enc2) -- (comp2);
            \draw[arrow] (enc2b) -- (comp2);
            \draw[arrow] (comp2) -- (dec2);
            \draw[arrow] (dec2) -- (output2);
            
            \node[above, font=\sffamily] at (0,4.5) {Audio Difference Captioning};
        \end{scope}

        \begin{scope}[xshift=12cm]
            \node (output3) at (0,4) {\small "Yes, there is a dog"};
            \node[box, fill=red!20] (dec3) at (0,2) {Answer Generator};
            \node[box, fill=gray!20] (fusion3) at (0,0) {Cross-Modal Fusion};
            \node[box, fill=blue!20] (enc3) at (-1.5,-2) {Audio Encoder};
            \node[box, fill=green!20] (qenc3) at (1.5,-2) {Question Encoder};
            \node (input3) at (-1.5,-4) {\includegraphics[width=1.5cm]{wijng1}};
            \node (question3) at (1.5,-4) {\small "Is there a dog?"};
            
            \draw[arrow] (input3) -- (enc3);
            \draw[arrow] (question3) -- (qenc3);
            \draw[arrow] (enc3) -- (fusion3);
            \draw[arrow] (qenc3) -- (fusion3);
            \draw[arrow] (fusion3) -- (dec3);
            \draw[arrow] (dec3) -- (output3);
            
            \node[above, font=\sffamily] at (0,4.5) {Audio Question Answering};
        \end{scope}

        \begin{scope}[xshift=18cm]
            \node (output4) at (0,4) {\small "dog barking in park"};
            \node[box, fill=purple!20] (sim4) at (0,2) {Similarity Search};
            \node[box, fill=gray!20] (db4) at (2,0) {Text Database};
            \node[box, fill=blue!20] (enc4) at (0,-2) {Audio Encoder};
            \node (input4) at (0,-4) {\includegraphics[width=1.8cm]{wijng1}};
            
            \draw[arrow] (input4) -- (enc4);
            \draw[arrow] (enc4) -- (sim4);
            \draw[arrow] (db4) -- (sim4);
            \draw[arrow] (sim4) -- (output4);
            
            \node[above, font=\sffamily] at (0,4.5) {Audio-to-Text Retrieval};
        \end{scope}

        \begin{scope}[xshift=0cm, yshift=-10cm]
            \node (output5) at (0,4) {\includegraphics[width=1.8cm]{wijng1}};
            \node[box, fill=orange!20] (voc5) at (0,2) {Vocoder};
            \node[box, fill=yellow!20] (dec5) at (0,0) {Audio Decoder};
            \node[box, fill=blue!20] (enc5) at (0,-2) {Text Encoder};
            \node (input5) at (0,-4) {\small "dog barking outside"};
            
            \node (noise5) at (2.5,0) {$\mathcal{N}(0,1)$};
            
            \draw[arrow] (input5) -- (enc5);
            \draw[arrow] (enc5) -- (dec5);
            \draw[arrow] (noise5) -- (dec5);
            \draw[arrow] (dec5) -- (voc5);
            \draw[arrow] (voc5) -- (output5);
            
            \node[above, font=\sffamily] at (0,4.5) {Audio Generation};
        \end{scope}

        \begin{scope}[xshift=6cm, yshift=-10cm]
            \node (output6) at (0,4) {\includegraphics[width=1.8cm]{wijng1}};
            \node[box, fill=purple!20] (sim6) at (0,2) {Similarity Search};
            \node[box, fill=gray!20] (db6) at (2,0) {Audio Database};
            \node[box, fill=blue!20] (enc6) at (0,-2) {Text Encoder};
            \node (input6) at (0,-4) {\small "dog barking outside"};
            
            \draw[arrow] (input6) -- (enc6);
            \draw[arrow] (enc6) -- (sim6);
            \draw[arrow] (db6) -- (sim6);
            \draw[arrow] (sim6) -- (output6);
            
            \node[above, font=\sffamily] at (0,4.5) {Text-to-Audio Retrieval};
        \end{scope}

        \begin{scope}[xshift=12cm, yshift=-10cm]
            \node (output7) at (0,4) {\includegraphics[width=1.8cm]{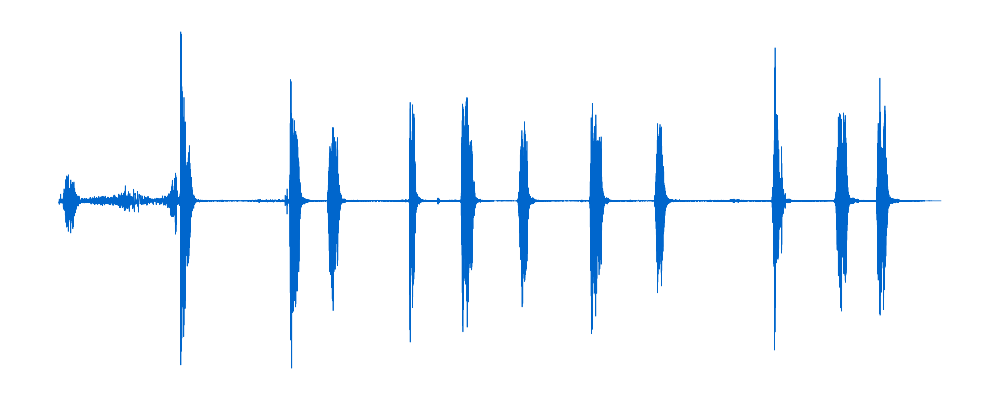}};
            \node[box, fill=red!20] (sep7) at (0,2) {Source Separator};
            \node[box, fill=gray!20] (fusion7) at (0,0) {Query-Audio Fusion};
            \node[box, fill=blue!20] (enc7) at (-1.5,-2) {Audio Encoder};
            \node[box, fill=green!20] (qenc7) at (1.5,-2) {Query Encoder};
            \node (input7) at (-1.5,-4) {\includegraphics[width=1.5cm]{wijng1}};
            \node (query7) at (1.5,-4) {\small "extract dog bark"};
            
            \draw[arrow] (input7) -- (enc7);
            \draw[arrow] (query7) -- (qenc7);
            \draw[arrow] (enc7) -- (fusion7);
            \draw[arrow] (qenc7) -- (fusion7);
            \draw[arrow] (fusion7) -- (sep7);
            \draw[arrow] (sep7) -- (output7);
            
            \node[above, font=\sffamily] at (0,4.5) {Audio Source Separation};
        \end{scope}

        \begin{scope}[xshift=18cm, yshift=-10cm]
            \node (output8) at (0,4) {\includegraphics[width=1.8cm]{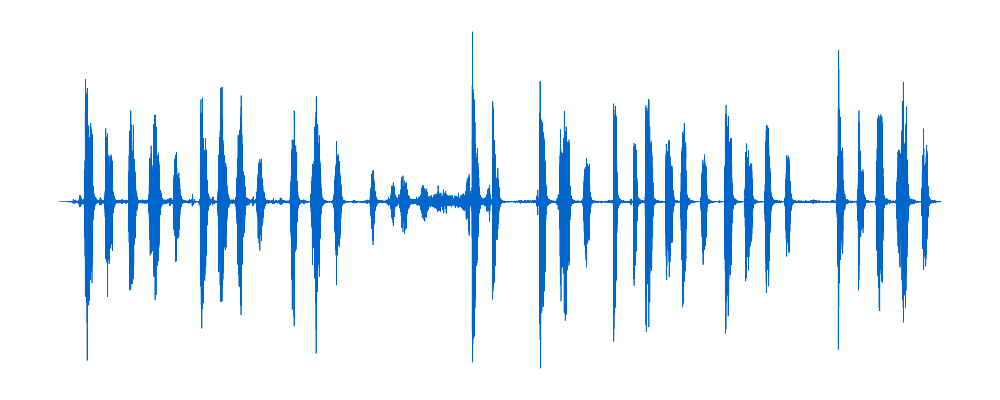}};
            \node[box, fill=orange!20] (edit8) at (0,2) {Audio Editor};
            \node[box, fill=gray!20] (fusion8) at (0,0) {Edit Fusion};
            \node[box, fill=blue!20] (enc8) at (-1.5,-2) {Audio Encoder};
            \node[box, fill=green!20] (inst8) at (1.5,-2) {Instruction Encoder};
            \node (input8) at (-1.5,-4) {\includegraphics[width=1.5cm]{wijng1}};
            \node (instr8) at (1.5,-4) {\small "add echo effect"};
            
            \draw[arrow] (input8) -- (enc8);
            \draw[arrow] (instr8) -- (inst8);
            \draw[arrow] (enc8) -- (fusion8);
            \draw[arrow] (inst8) -- (fusion8);
            \draw[arrow] (fusion8) -- (edit8);
            \draw[arrow] (edit8) -- (output8);
            
            \node[above, font=\sffamily] at (0,4.5) {Audio Editing};
        \end{scope}
    \end{tikzpicture}
    }
    \label{fig:audio_language_models}
\end{figure*}

\section{Background} \label{sec:background}
This section provides an introductory overview of Audio-Language Learning, categorizing the field into two primary task types: \textbf{Audio-to-Language} and \textbf{Language-to-Audio}. The key tasks within these categories are outlined and illustrated in Figure \ref{fig:audio_language_models}. Readers interested in in-depth descriptions of specific tasks are directed to the surveys by \citet{xuStatusQuoContemporary2024} and \citet{wangCrossModalRetrievalSystematic2024}.

\subsection{Audio-to-Language}
The Audio-to-Language category comprises models that transform audio inputs into textual outputs, either by generating text or retrieving it from a dataset. This category encompasses several subfields:
\begin{enumerate}
\item \textbf{Audio Captioning}: Generating natural language descriptions for audio events. 
\item \textbf{Audio Difference Captioning}: Highlighting content differences between pairs of similar audio clips (e.g. the addition or removal of specific sounds). 
\item \textbf{Audio Representation Learning}: Learning audio representations that are effective for downstream tasks.
\item \textbf{Audio-Text Retrieval}: Using a database of audio-language pairs to find textual descriptions corresponding to given audio inputs.
\item \textbf{Audio-Question Answering}: Answering natural language questions about specific audio content.

\end{enumerate}

\subsubsection{(Automated) Audio Captioning (AAC)} \label{sec:audiocaptioning}
As introduced by \citet{drossosAutomatedAudioCaptioning2017}, AAC refers to converting audio signals into natural language descriptions. Prominent datasets within this domain include Clotho \cite{drossosClothoAudioCaptioning2020} and AudioCaps \cite{kimAudioCapsGeneratingCaptions2019} (see Section \ref{sec:datasets}). AAC models predominantly utilize an encoder-decoder framework. The \textbf{encoder} processes audio signals into latent representations. The \textbf{decoder} converts these representations into textual descriptions. Encoders are frequently pre-trained on datasets such as AudioSet \cite{gemmekeAudioSetOntology2017} for audio classification tasks \cite{kohAutomatedAudioCaptioning2021, meiAudioCaptioningTransformer2021}, facilitating transfer learning to AAC tasks, as demonstrated in various studies \cite{wonTransferLearningFollowed2021, meiEncoderDecoderBasedAudio2021, kadlcikWhisperTransformerAudio2023}. Common encoders architectures include Convolutional Neural Networks (CNNs) \cite{gemmekeAudioSetOntology2017, kongPANNsLargeScalePretrained2020} and Transformer encoders \cite{gongASTAudioSpectrogram2021}. Decoders typically employ Transformers \cite{radfordLanguageModelsAre2019, lewisBARTDenoisingSequenceSequence2019}. Transformers and RNNs allow for variable-length audio input, while CNNs are often limited to fixed-length input due to their pre-defined kernel size and spatial dimensions. Various other architectures have been proposed, such as GANs \cite{meiDiverseAudioCaptioning2022a}, diffusion models \cite{xuDiverseEfficientAudio2024}, and variational audioencoders \cite{zhangGeneratingAccurateDiverse2024}.

AAC models are trained on paired audio-text data $(\tau, \alpha)$, employing various training methods. Early models used encoder-decoder architectures, where an audio encoder produced embeddings for a text decoder. Contemporary approaches include cross-attention mechanisms enabling multimodal audio-language fusion \cite{gontierAutomatedAudioCaptioning2021, koizumiAudioCaptioningUsing2020a,chenAudioCaptioningBased2020, meiEncoderDecoderBasedAudio2021}, integration using audio embeddings as prefixes for text decoders \cite{schaumloffelPEACSPrefixEncoding2023, kimPrefixTuningAutomated2023}, and zero-shot approaches adapting pre-trained multimodal models and LLMs without direct training \cite{shaharabanyZeroShotAudioCaptioning2023, salewskiZeroshotAudioCaptioning2023, miaoZeroshotTransferWildlife2023}. Innovative variations include retrieval-augmented captioning \cite{ghoshRecapRetrievalaugmentedAudio2024} and unified encoder architectures \cite{chuQwenAudioAdvancingUniversal2023, tangSALMONNGenericHearing2024}, which allow for sophisticated integration of audio and text modalities.

A powerful training technique that has emerged for audio captioning models is Instruction Tuning, which improves model performance with reformatted training data of instruction-response pairs. This approach has demonstrated superior results compared to traditional training techniques \cite{daiInstructBLIPGeneralpurposeVisionLanguage2023}. Two primary strategies have been explored: using manually crafted templates \cite{weiFinetunedLanguageModels2022, chungScalingInstructionFinetunedLanguage2024, sanhMultitaskPromptedTraining2022, wangSuperNaturalInstructionsGeneralizationDeclarative2022} or leveraging LLMs to generate instruction datasets automatically \cite{honovichUnnaturalInstructionsTuning2023, wangSelfInstructAligningLanguage2023}. Notable examples include Vicuna \cite{zhengJudgingLLMJudgeMTBench2023}, Stanford Alpaca \cite{rohantaoriAlpacaStrongReplicable}, and LLaMA-GPT4 \cite{pengInstructionTuningGPT42023} all developed by fine-tuning LLaMA using LLM-generated instruction datasets. This technique has also proven effective in multimodal contexts, such as guiding visual encoders to extract and explain image features using instructions \cite{zhuMiniGPT4EnhancingVisionLanguage2024}.

The task of Audio Logging (AL) extends the concept of AAC by generating longer, paragraph-length descriptions of audio content. For example, \citet{baiAudioLogLLMsPoweredLong2023} introduced a method that uses LLMs to describe audio recordings up to three minutes in duration.

The task of Soundscape Captioning, as proposed by \cite{houSoundscapeCaptioningUsing2024a}, aims to generate context-aware descriptions of soundscapes by capturing the acoustic scene, event information, and corresponding human affective qualities. This task seeks to automate traditional soundscape analysis, which relies on subjective ratings and labor-intensive surveys.

Audio Entailment, introduced by \cite{deshmukhAudioEntailmentAssessing2024}, involves evaluating whether a text description can be inferred from an audio recording. Each text description is categorized as either an entailment, neutral or contradiction with respect to the audio. This task assesses an audio-language model's deductive reasoning capabilities by determining whether a text hypothesis can be logically inferred from an audio premise.

Text-queried Sound Event Localization (SEL), proposed by \cite{zhaoTextqueriedTargetSound2024}, involves identifying the location or direction of a specific sound event described in natural language. Unlike traditional Sound Event Localization and Detection (SELD) systems, which are limited to predefined classes, SEL models allow users to describe target sounds in natural language, making the task more flexible and user-centric.

\subsubsection{Audio Difference Captioning (ADC)}
Audio Difference Captioning (ADC) addresses a limitation in traditional audio captioning, where models often generate similar descriptions for similar sounds without adequately capturing the differences in their content. ADC models utilize annotations that explicitly highlight differences between captions of similar audio clips, enabling better differentiation. For example, \citet{tsubakiAudiochangeCaptioningExplain2023} demonstrated the application of ADC in detecting small anomalies in machine sounds to showcase its usefulness in real-world scenarios. Prominent datasets for ADC include: (1) the MIMII-Change dataset \cite{tsubakiAudiochangeCaptioningExplain2023}, which focuses on linguistic descriptions of anomalies in machine-sounds. (2) The AudioDiffCaps dataset \cite{takeuchiAudioDifferenceCaptioning2023}, which features audio clips with artificially mixed background and foreground sounds derived from existing environmental sound datasets.

A related technique, termed \textit{Audio Difference Learning}, is described by \citet{komatsuAudioDifferenceLearning2023}, specifically as a pre-training strategy for Audio Captioning models. In this approach, two audio inputs - an input and a reference audio - are mixed and encoded. The encoded reference is subtracted from the encoded input, and the result is passed to a decoder to generate the caption. This method bypasses the need to create specialised ADC datasets, streamlining the training process. 

Additionally, a closely related task called \textit{Natural Language Audio Reasoning} has been proposed by \cite{liangAcousticPromptTuning2023}. NLAR focuses on distinguishing, comparing and summarizing differences between two audio clips. To support this task, the researchers introduced the NLAR dataset, which has proven effective in evaluating the reasoning capabilities of audio models.

\subsubsection{Audio Representation Learning (ARL)}
Audio Representation Learning (ARL) focuses on learning audio representations useful for downstream tasks. This typically involves training a mapping function $f$ that transforms an audio clip $x$ into a meaningful representation $y$. The mapping function $f$ is often learned using techniques like contrastive learning \cite{wuWav2CLIPLearningRobust2022} or self-supervised learning \cite{chenAudioCaptioningBased2020, meiEncoderDecoderBasedAudio2021}. These methods aim to create robust embeddings that capture essential features of audio signals, enabling enhanced performance across various applications.

\subsubsection{Audio-to-Text Retrieval (ATR)}
Audio-to-Text Retrieval (ATR) involves finding the textual descriptions corresponding to a given audio query. This task is challenging due to the inherent modality gap between audio and text modalities, requiring models are able to align these distinct data types. Recent research has emphasized developing cross-modal alignment techniques to bridge this gap effectively \cite{meiMetricLearningAudioText2022, luoContrastiveLatentSpace2023}.

Contrastive learning has emerged as a key approach in ATR. It trains models to maximize similarity between paired audio-text samples while minimizing similarity with unrelated pairs. For example, \citet{wuTextaudioRetrievalLargescale2022} employed a CLIP-inspired contrastive learning framework to align audio and text embeddings in a shared space. Similarly, \citet{deshmukhAudioRetrievalWavtext5k2022} introduced a framework that combines two audio encoders with a text encoder and contrastive learning to enhance retrieval performance for variable-length audio content.

\subsubsection{Audio-Question Answering (AQA)}
Audio-Question Answering (AQA) is a multimodal task where a system receives audio input and a related textual question to generate accurate answers about the audio content. This requires integrating audio signal processing with natural language understanding to produce contextually relevant responses. Advances in this field have been driven by attention-based neural network models that utilize self-attention to create representations for both audio and text. Cross-attention mechanisms map relevant audio features to textual queries, improving precision \cite{liMultiscaleAttentionAudio2023, sudarsanamAttentionBasedMethodsAudio2023}.

AQA datasets differ from typical audio-language datasets using a triplet structure: an audio file, a text-based question, and a corresponding text-based answer. This setup supports interactive models beyond understanding or transcribing audio to interpreting and answering specific queries. The question and answer components are textual, with answers often structured for classification tasks, such as simple yes/no responses. For analytical purposes, questions and answers are sometimes concatenated into a single text string (see Table \ref{tab:datasets}).

A sub-task of AQA is Spatial Sound Question Answering, introduced by \citet{zhengBATLearningReason2024}. This variation aims to spatially detect and localize sounds by determining their direction of arrival and distance, adding a spatial reasoning dimension to the task.

\subsection{Language-to-Audio}
The Language-to-Audio category encompasses tasks where language input generates, retrieves or modifies audio.  
This category is divided into four key fields: 
\begin{enumerate}
\item \textbf{Audio Generation}: The creation of audio based on natural language prompts.
\item \textbf{Text-To-Audio Retrieval}: Retrieving audio clips from a database using natural language queries. 
\item \textbf{Audio Editing}: The modification of existing audio content based on language instructions.
\item \textbf{Language-Queried Audio Source Separation}: The isolation of specific sounds from a mixture using natural language queries.
\end{enumerate}

\subsubsection{Audio Generation (AG)} 
Audio generation, also called Audio Synthesis or Text-To-Audio (TTA), focuses on producing synthetic audio that can mimic sounds, speech, or music. These models accept text inputs (prompts) and generate diverse audio outputs. The main approaches to audio generation include:
\begin{enumerate}
	\item Language-model based methods: Autoregressive models trained on text-audio pairs, such as AudioGen \cite{kreukAudioGenTextuallyGuided2023} and AudioLM \cite{borsosAudioLMLanguageModeling2023} generate audio directly from textual prompts.
	\item Diffusion-based models: Latent diffusion models, like Diffsound \cite{yangDiffsoundDiscreteDiffusion2023} and AudioLDM \cite{liuAudioLDMTexttoAudioGeneration2023} generate audio from scratch or conditionally based on input text or audio. These have become the dominant approach in audio generation \cite{baiConsistencyTTAAcceleratingDiffusionBased2024a}.
	\item Consistency models: Models such as AudioLCM \cite{liuAudioLCMTextAudioGeneration2024} synthesize audio with high efficiency, requiring only two iterations. 
\end{enumerate}

\subsubsection{Text-to-Audio Retrieval (TAR)}
Text-to-audio retrieval (TAR), or language-based audio retrieval, involves searching for and retrieving audio clips that match a given text query \cite{oncescuAudioRetrievalNatural2021}. This task parallels Audio-to-Text Retrieval, focusing on mapping audio and text into a shared embedding space for cross-modal retrieval. Approaches include contrastive learning \cite{wuTextaudioRetrievalLargescale2022, laiResnetbasedClipTextaudio2022}, and employing training objectives such as triplet loss \cite{xieLanguagebasedAudioRetrieval2022, pellegriniLanguagebasedAudioRetrieval2022}.
Subtasks include Audio-Text Relevance Learning, where the goal is to learn the shared semantic properties of audio samples and text description, using similarity functions \cite{xieIntegratingContinuousBinary2024}, and Audio Moment Retrieval \cite{munakataLanguagebasedAudioMoment2024}, where the models identify specific timestamps in audio that correspond to a text query.

\subsubsection{Audio Editing}
Audio editing focuses on altering existing audio content based on language instructions. Recent advancements include training-free methods, like AudioEditor \cite{jiaAudioEditorTrainingfreeDiffusionbased2024}, which repurpose pre-trained text-to-audio models using techniques such as null-text inversion and EOT suppression, and specifically trained models:  like AUDIT \cite{wangAUDITAudioEditing2023}, which employ triplet training data and targeted modification learning.
These methods enable tasks like adding background sounds, replacing instruments, and repairing audio while maintaining the integrity of unedited sections.

\subsubsection{Language-queried audio source separation (LASS)} \label{sec:lass}
Language-queried audio source separation (LASS) uses natural language descriptions to isolate specific sound sources from a mixture of sounds. 
This field is also referred to as Text-to-Audio Grounding \cite{xuTextAudioGroundingBuilding2021, xuWeaklySupervisedTextAudio2024}, Language-assisted Sound Event Detection (SED-LM) \cite{wangLeveragingLanguageModel2024}, Language-Queried Target Sound Extraction \cite{xuLeveragingSoundLocal2023}, Query-based Sound Separation (QSS) \cite{liuSeparateAnythingYou2023a}, Query-conditioned target sound extraction \cite{maCLAPSepLeveragingContrastive2024}.

The prominent models in this field include LASSNet \cite{liuSeparateWhatYou2022}, SoundWords \cite{kilgourTextdrivenSeparationArbitrary2022}, CLIPSep \cite{dongCLIPSepLearningTextqueried2023}, CLAPSep \cite{maCLAPSepLeveragingContrastive2024}, AudioSep \cite{liuSeparateAnythingYou2023a} and FlowSep \cite{yuanFlowSepLanguagequeriedSound2024}. In 2024, LASS was introduced as a new challenge in the DCASE competition; the winner was AudioSep-DP \cite{yinExploringTextqueriedSound2024}. 
\begin{figure*}[p]
    \caption{This figure indicates various datasets and which task in the field of Audio-Language Learning the dataset was originally proposed for. The color indicates if the dataset overlaps with another dataset, based on their origin. The background of datasets that consists of multiple colors overlap in multiple origin datasets. The derivative of Clotho are placed under Freesound, and derivatives of AudioCaps under YouTube. Datasets that do not share origin with other datasets are marked with a white background. \textsuperscript{\textdagger} = datasets for text-to-audio retrieval are synonymous to datasets for audio-to-text retrieval. }

    \label{fig:literature_tree}
	\centering
        \pgfdeclarehorizontalshading{youtube_freesound}{100bp}{
   color(0bp)=(orange!20);
   color(45bp)=(orange!20);
   color(55bp)=(purple!20);
   color(100bp)=(purple!20)
        } 
        \pgfdeclarehorizontalshading{bbc_freesound}{100bp}{
   color(0bp)=(blue!20);
   color(45bp)=(blue!20);
   color(55bp)=(purple!20);
        	color(100bp)=(purple!20)
}
\pgfdeclarehorizontalshading{youtube_prosoundeffects}{100bp}{
   color(0bp)=(orange!20);
   color(45bp)=(orange!20);
   color(55bp)=(cyan!20);
        	color(100bp)=(cyan!20)
}
		\pgfdeclarehorizontalshading{youtube_bbc_freesound}{100bp}{
        	color(0bp)=(orange!20);
   color(35bp)=(orange!20);
   color(40bp)=(blue!20);
        	color(60bp)=(blue!20);
   color(65bp)=(purple!20);
        	color(100bp)=(purple!20)
} 
		\pgfdeclarehorizontalshading{soundbible_bbc_freesound}{100bp}{
        	color(0bp)=(teal!20);
   color(35bp)=(teal!20);
   color(40bp)=(blue!20);
        	color(60bp)=(blue!20);
   color(65bp)=(purple!20);
        	color(100bp)=(purple!20)
} 

\pgfdeclarehorizontalshading{youtube_macs_freesound}{100bp}{
        	color(0bp)=(orange!20);
   color(35bp)=(orange!20);
   color(40bp)=(green!20);
        	color(60bp)=(green!20);
   color(65bp)=(purple!20);
        	color(100bp)=(purple!20)
} 
\pgfdeclarehorizontalshading{youtube_soundbible_freesound}{100bp}{
        	color(0bp)=(orange!20);
   color(35bp)=(orange!20);
   color(40bp)=(teal!20);
        	color(60bp)=(teal!20);
   color(65bp)=(purple!20);
        	color(100bp)=(purple!20)
} 
\pgfdeclarehorizontalshading{youtube_bbc_macs_freesound}{100bp}{
        	color(0bp)=(orange!20);
   color(25bp)=(orange!20);
   color(30bp)=(blue!20);
        	color(50bp)=(blue!20);
   color(55bp)=(green!20);
   color(75bp)=(green!20);
   color(80bp)=(purple!20);
        	color(100bp)=(purple!20)
} 
\pgfdeclarehorizontalshading{youtube_bbc_soundbible_freesound}{100bp}{
        	color(0bp)=(orange!20);
   color(25bp)=(orange!20);
   color(30bp)=(blue!20);
   color(50bp)=(blue!20);
   color(55bp)=(teal!20);
   color(75bp)=(teal!20);
   color(80bp)=(purple!20);
        	color(100bp)=(purple!20)
} 
        \tikzset{
   basic/.style  = {draw, rectangle, text width=2.5cm, align=center, font=\sffamily\tiny, inner sep=1pt},
   rootspace/.style = {font=\sffamily\tiny, text width=2.7cm, draw=none},
   root/.style   = {font=\sffamily\tiny}, 
   onode/.style = {basic, thin, rounded corners=2pt, align=center, fill=green!60, text width=2.5cm},
   tnode_self/.style = {basic, thin, align=left, text width=6em, align=center},
   tnode_macs/.style = {basic, thin, align=left, fill=green!20, text width=6em, align=center},
   tnode_youtube/.style = {basic, thin, align=left, fill=orange!20, text width=6em, align=center},
   tnode_bbc/.style = {basic, thin, align=left, fill=blue!20, text width=6em, align=center},
   tnode_freesound/.style = {basic, thin, align=left, fill=purple!20, text width=6em, align=center},
   tnode_soundbible/.style = {basic, thin, align=left, fill=teal!20, text width=6em, align=center},
   tnode_prosoundeffects/.style = {basic, thin, align=left, fill=cyan!20, text width=6em, align=center},
   tnode_yf/.style = {basic, thin, align=left, shading=youtube_freesound, shading angle=0, text width=6em, align=center},
   tnode_bf/.style = {basic, thin, align=left, shading=bbc_freesound, shading angle=0, text width=6em, align=center},
   tnode_yp/.style = {basic, thin, align=left, shading=youtube_prosoundeffects, shading angle=0, text width=6em, align=center},
   tnode_sbf/.style = {basic, thin, align=left, shading=soundbible_bbc_freesound, shading angle=0, text width=6em, align=center},
   tnode_ybf/.style = {basic, thin, align=left, shading=youtube_bbc_freesound, shading angle=0, text width=6em, align=center},
   tnode_ysf/.style = {basic, thin, align=left, shading=youtube_soundbible_freesound, shading angle=0, text width=6em, align=center},
   tnode_ymf/.style = {basic, thin, align=left, shading=youtube_macs_freesound, shading angle=0, text width=6em, align=center},
   tnode_ybmf/.style = {basic, thin, align=left, shading=youtube_bbc_macs_freesound, shading angle=0, text width=6em, align=center},
   tnode_ybsf/.style = {basic, thin, align=left, shading=youtube_bbc_soundbible_freesound, shading angle=0, text width=6em, align=center},
   edge from parent/.style={draw=black, edge from parent fork right},
        }

        \begin{forest}
        for tree={
   grow=east,
   growth parent anchor=west,
   parent anchor=east,
   child anchor=west,
   edge path={
\noexpand\path[\forestoption{edge},->, >={latex}] 
(!u.parent anchor) -- +(5pt,0pt) |- (.child anchor)
\forestoption{edge label};
   },
   l sep=8mm,
   s sep=1mm, 
        }
        [Audio-Language Datasets for Sounds and Events, basic,
   [Language-to-Audio, root,
[Language-Queried Audio Source Separation, root,
    [CAPTDURE \cite{okamotoCAPTDURECaptionedSound2023}, tnode_self] 
    [LASS (DCASE-Synth \& DCASE-Real) \cite{liuSeparateWhatYou2022, liuSeparateAnythingYou2023a}, tnode_freesound]
    [AudioGrounding (v2) \cite{xuTextAudioGroundingBuilding2021}, tnode_youtube]
]
[Audio Editing, root]
[Text-to-Audio Retrieval, root,
    [Audio-Text Relevance Learning, root,
 [Audio-Alpaca \cite{majumderTango2Aligning2024}, tnode_youtube]
 [Syncaps \cite{oncescuDissectingTemporalUnderstanding2024a}, tnode_freesound] 
    ]
    [Audio Moment Retrieval, root,  
 [Clotho-Moment \cite{munakataLanguagebasedAudioMoment2024}, tnode_freesound]
    ]
    [Standard Text-to-Audio Retrieval \textsuperscript{\textdagger}, root,
 [AnimalSpeak \cite{robinsonTransferableModelsBioacoustics2024}, tnode_youtube] 
 [AudioEgoVLP \cite{oncescuSOUNDAPPROACHUsing2024a}, tnode_self] 
 [ClothoV2\_GPT \cite{primusAdvancingNaturalLanguageBased2023}, tnode_freesound]
 [SoundDescs \cite{koepkeAudioRetrievalNatural2022}, tnode_bbc] 
 [SoundingEarth \cite{heidlerSelfsupervisedAudiovisualRepresentation2023}, tnode_self] 
 [WavText5K \cite{deshmukhAudioRetrievalWavtext5k2022}, tnode_soundbible] 
    ]
]
[Audio Generation, root,
    [AFAudioSet \cite{kongImprovingTextAudioModels2024}, tnode_youtube] 
    [AudioCondition \cite{guoAudioGenerationMultiple2024}, tnode_youtube] 
    [AudioTPE \cite{wangAudioComposerFinegrainedAudio2024}, tnode_freesound] 
    [AudioTime \cite{xieAudioTimeTemporallyalignedAudiotext2024}, tnode_youtube] 
    [BATON \cite{liaoBATONAligningTexttoaudio2024}, tnode_youtube] 
    [PicoAudio \cite{xiePicoaudioEnablingPrecise2024}, tnode_freesound] 
    [RichDetail-AudioTextSimulation \cite{xuDetailedAudiotextData2024}, tnode_freesound] 
    [Sound-VECaps\_A \cite{yuanSoundVECapsImprovingAudio2024}, tnode_youtube] 
    [TangoPromptBank \cite{ghosalTextaudioGenerationUsing2023}, tnode_ybsf] 
    [VGGSound-Caption \cite{yangDrawAudioLeveraging2024}, tnode_youtube] 
]
   ]
   [Audio-to-Language, root,
[Audio Question Answering, root,
    [Standard Audio Question Answering, root,
 [AQUALLM \cite{beheraAQUALLMAudioQuestion2023a}, tnode_ymf] 
 [Audio Dialogues \cite{goelAudioDialoguesDialogues2024a}, tnode_youtube] 
 [ClothoAQA \cite{lippingClothoAQACrowdsourcedDataset2022}, tnode_freesound] 
 [DAQA \cite{fayekTemporalReasoningAudio2020}, tnode_youtube] 
 [LTU / OpenAQA-5M \cite{gongJointAudioSpeech2023}, tnode_ysf] 
 [mClothoAQA \cite{beheraMultiLingualAudioQuestion2023}, tnode_freesound] 
 [MULTIS \cite{zhaoChatBridgeBridgingModalities2023}, tnode_yf] 
    ]
    [Spatial Sound Question Answering, root,
 [SpatialSoundQA \cite{zhengBATLearningReason2024}, tnode_youtube] 
    ]
]
[Audio Representation Learning, root,
    [ACalt4 \cite{niizumiM2DCLAPMaskedModeling2024}, tnode_youtube]
    [LAION-Audio-630k \cite{wuLargeScaleContrastiveLanguageAudio2023}, tnode_bf] 
]
[Audio Difference Captioning, root,
    [AudioDiffCaps \cite{takeuchiAudioDifferenceCaptioning2023}, tnode_freesound]
    [MIMII-Change \cite{tsubakiAudiochangeCaptioningExplain2023}, tnode_self] 
    [NLAR \cite{liangAcousticPromptTuning2023}, tnode_freesound]
]
[Audio Captioning, root,
    [Standard Audio Captioning, rootspace,
 [AS-Qwen-Caps \& AS-SL-GPT4-Caps \cite{haiEzAudioEnhancingTextaudio2024}, tnode_youtube] 
 [AudioCaption \cite{wuAudioCaptionListen2019, xuAudioCaptionCar2021}, tnode_self]
 [AudioCaps \cite{kimAudioCapsGeneratingCaptions2019}, tnode_youtube]
 [AudioHallucination \cite{kuanUnderstandingSoundsMissing2024}, tnode_youtube]
 [AudioSetMix \cite{xuAudioSetMixEnhancingAudioLanguage2024a}, tnode_youtube]
 [Auto-ACD \cite{sunAutoACDLargescaleDataset2024}, tnode_youtube]
 [AutoCap Dataset \cite{haji-aliTamingDataTransformers2024a}, tnode_ybsf] 
 [CompA-R \cite{ghoshGAMALargeAudioLanguage2024a}, tnode_youtube] 
 [Clotho \cite{drossosClothoAudioCaptioning2020}, tnode_freesound]
 [Clotho-ChatGPT-Mixup \cite{wuImprovingAudioCaptioning2024}, tnode_freesound]
 [Clotho-Detail \cite{zhaoBuboGPTEnablingVisual2023}, tnode_freesound]
 [FAVDBench Audio \cite{shenFinegrainedAudibleVideo2023}, tnode_self]
 [Freesound 500K \cite{tangAnyAnyGenerationComposable2023a}, tnode_freesound]
 [MACS \cite{martin-moratoWhatGroundTruth2021, martinmoratoDiversityBiasAudio2021}, tnode_macs] 
 [Pro Sound Effects Library \cite{priceProSoundEffects}, tnode_prosoundeffects]
 [Revised-Clotho \cite{xiaoEnsembleSystemsContrastive2023}, tnode_freesound]
 [SAM-S \cite{hebbarDatasetAudioVisualSound2023}, tnode_self]
 [WavCaps \cite{meiWavCapsChatGPTAssistedWeaklyLabelled2023}, tnode_sbf]
    ]
    [Text-queried Sound Event Localization, root]
    [Audio Entailment, root,
 [AudioCaps \& Clotho Entailment \cite{deshmukhAudioEntailmentAssessing2024}, tnode_yf]
    ]
    [Soundscape Captioning, root,
 [SoundScaper Dataset \cite{houSoundscapeCaptioningUsing2024a}, tnode_self] 
    ]
    [Audio Logging, root,
 [AudioLog Dataset \cite{baiAudioLogLLMsPoweredLong2023}, tnode_self]  
    ]
]
   ]
        ]
        \end{forest}
        \begin{tikzpicture}[remember picture, overlay]
   \tikzset{
legendbox/.style={
    draw,
    minimum size=1em,
    inner sep=0pt,
    anchor=base,
    yshift=0.5ex
},
legendtext/.style={
    thin,
    basic,
    draw=none,
    anchor=base west,
    align=left
}
   }
   \node[anchor=north west] at ([xshift=2cm, yshift=-6cm]current page.north west) {
   \begin{tikzpicture}[font=\tiny]
\matrix[
    draw,
    fill=white,
    inner sep=4pt,
    row sep=2pt,
    column sep=2pt,
    nodes={
 font=\small
    }
] {
    \node[fill=orange!20, legendbox] {}; & \node[legendtext] {YouTube}; \\
    \node[fill=purple!20, legendbox] {}; & \node[legendtext] {Freesound}; \\
    \node[fill=blue!20, legendbox] {}; & \node[legendtext] {BBC}; \\
    \node[fill=teal!20, legendbox] {}; & \node[legendtext] {SoundBible}; \\
    \node[fill=green!20, legendbox] {}; & \node[legendtext] {MACS}; \\
    \node[fill=cyan!20, legendbox] {}; & \node[legendtext] {Pro Sound Effects}; \\
};
   \end{tikzpicture}
   };
        \end{tikzpicture}
\end{figure*}

\section{Audio-language datasets} \label{sec:datasets}
Datasets play an important role in training models in audio-language tasks (see Figure \ref{fig:literature_tree}). However, most existing audio datasets are relatively small and domain-specific \cite{fonsecaFSD50KOpenDataset2022, sunAutoACDLargescaleDataset2024}, limiting their utility in broader applications. Annotating these datasets often omits background details or common sense relationships, as authors focus primarily on the most salient audio features \cite{betkerImprovingImageGeneration2023}. In addition, researchers face significant challenges in the time-intensive process of collecting and curating high-quality audio-language datasets \cite{meiWavCapsChatGPTAssistedWeaklyLabelled2023}.

Several datasets used in the training of audio models, particularly those sourced from platforms like YouTube, exhibit varying levels of noise and reliability \cite{liRobustLargescaleAudio2023}. Clips from these platforms are often recorded with consumer devices in real-life settings and may include diverse background sounds, such as human speech, music, and environmental noises (e.g., wind), which are frequently omitted from captions. The trade-off when using such datasets lies in balancing size and quality. While larger datasets are easier to compile with noisier and less curated content, this approach does not necessarily translate into improved model performance\cite{palm2teamPaLM2Technical2023}.

Another common practice in audio-language models involves exploiting datasets originally designed for visual tasks (e.g., \cite{chenVGGSoundLargescaleAudioVisual2020, sunAutoACDLargescaleDataset2024}). However, these visual-centric datasets often include captions that describe visually observable details irrelevant to the audio, such as "the man is wearing a blue sweater" or color and shapes of objects \cite{kulikTakeItGranted2024}. Additionally, sounds in videos may originate from off-screen events \cite{xuBLATBootstrappingLanguageAudio2023}. Such discrepancies can lead to bias during training, where models may learn to predict irrelevant visual attributes instead of audio-related ones. 

For this reason, visual-centric datasets are excluded from the main analysis in this survey. Nonetheless, they may still be valuable for training in scenarios where visual descriptions complement the intended use case, and they are included in the Appendix \ref{app:audio-visual-datasets}.

This section first highlights how this survey differentiates itself from recent work. Next, it describes the methods used to identify datasets relevant to Audio-Language Learning. Finally, it provides detailed descriptions of the identified dataset, including their key features and applications.

\subsection{Related work} \label{sec:related-work}
To the best of our knowledge, six previous surveys examined the audio-language domain. Three surveys focus on AAC \cite{meiAutomatedAudioCaptioning2022, xuStatusQuoContemporary2024, nagSystematicLiteratureReview2021}, one examines audio retrieval \cite{wangCrossModalRetrievalSystematic2024}, another explores the use of LLM in audio processing \cite{latifSparksLargeAudio2023a}, and the final one addresses traditional and foundation models in sound understanding \cite{triantafyllopoulosComputerAuditionTaskspecific2024}.

\citet{meiAutomatedAudioCaptioning2022} provide insights into fundamental concepts, as well as training and evaluation methodologies in AAC. Similarly, \citet{xuStatusQuoContemporary2024} focus on AAC, specifically on dataset augmentation and models with additional guidance. \citet{xuStatusQuoContemporary2024} also highlights future directions, such as controllable and diverse captioning. Both surveys especially tackle training techniques and supplementary strategies for improvements. 

\citet{nagSystematicLiteratureReview2021} is a thesis reporting a literature review on visual and audio captioning, describing existing methods, models, popular datasets, and evaluation metrics, with a particular emphasis on captioning techniques.

\citet{wangCrossModalRetrievalSystematic2024} offer a review of cross-modal retrieval, including a taxonomy of machine learning paradigms, mechanisms and models. They also discuss benchmarks, performance metrics and future research directions. \citet{latifSparksLargeAudio2023a} examine the application of LLMs in audio processing, including speech and music, reviewing recent developments and associated challenges. 

\citet{triantafyllopoulosComputerAuditionTaskspecific2024} provide an overview of foundational and traditional models for sound understanding tasks, such as Audio Classification and Sound Event Detection. It contains a comprehensive list of 20 datasets, but it lacks a detailed computational analysis of the datasets.

Our survey is inspired by the work of \cite{tavaresClassSeparabilityPitfalls2024}, highlighting the need for a deeper understanding of audio datasets and their relationships. The authors demonstrated that apparent zero-shot learning capabilities in audio-text models may originate from pre-trained backbone models rather than genuine cross-modal learning. This raises questions about the feasibility of true zero-shot learning, given the potential overlap between training and evaluation data. Our survey addresses these concerns by analyzing the relationships and differences between audio datasets, providing a foundation for understanding data leakage and overlap that may affect model evaluation. 

\subsection{Method}
The datasets selected for this survey were identified through a systematic manual search and review of publications indexed in Google Scholar from 2019 to 2023. The survey started by selecting publications that referenced at least one known task and one dataset within the audio-language field. Using the search term below in Google Scholar via Publish or Perish, a total of 409 papers were retrieved: ("Audio Language" OR  "Audio Captioning" OR "Audio-Text Retrieval" OR "Audio-Question Answering" OR "Language-to-Audio" OR "Audio Generation" OR "Text-to-Audio Retrieval") AND ("Clotho" OR "AudioCaps" OR "MACS" OR "Auto-ACD" OR "AudioCaption" OR "WavText5K" OR "SoundDescs" OR "WavCaps").

We thoroughly reviewed each of these 409 papers, examining them for mentions of datasets and key insights regarding audio-language learning. This included analyzing their discussions on evaluation methodologies, model architectures, and challenges in the field. We paid particular attention to papers that introduced new datasets or made significant observations about existing ones, especially regarding zero-shot learning capabilities, data quality and biases.

For an updated version of this survey, an additional search was conducted through Publish or Perish using the same search term, covering publications from January to October 1st 2024. This search yielded an additional 299 papers, which underwent the same rigorous review process. This comprehensive analysis consists of both paper sets and we identified and cataloged the datasets presented in Table \ref{tab:datasets}, along with their key characteristics and relationships to different audio-language tasks.

Please note that this survey incorporates research up to and including September 2024. Papers published after this period were not included, so the findings may not reflect the most recent advancements in the field.

\subsection{Datasets}
Audio captioning datasets can be categorized into several groups based on their sources:

\begin{itemize}
\item YouTube-Based: Includes datasets derived from YouTube, such as AudioSet and AudioCaps, which originates from AudioSet.
\item Freesound-Based: Clotho being a primary example, these are datasets from the Freesound database.
\item Sound-Effect Datasets: Examples include WavText5K and the BBC Sound Effects Library.
\item Field-recording Datasets: These datasets are sourced from field recordings, such as the MACS dataset. Field recordings are audio samples captured in natural environments, such as urban, rural, or natural settings.
\item Movies \& TV Shows Datasets: These datasets are sourced from television shows, such as the AudioCaption dataset.
\end{itemize}

Table \ref{tab:datasets} presents a summary of datasets relevant to the field of audio-language learning. This subsection contains a brief description of each dataset. The table in Appendix \ref{app:dog-captions} contains example captions for each dataset discussed here. Moreover, Table \ref{tab:alm-overview} illustrates the combinations of datasets employed to train large-scale audio language models.

\begin{table*}[ht!]
\caption{Comparative overview of 69 datasets in the audio-language domain. The statistics are self-calculated where possible. Indentation indicates reuse of the audio or that the dataset is a subset of another dataset. Use Case indicates what sets the dataset apart from others. Easily accessible datasets are publicly available and do not require special access, scraping, payment or complicated access procedures. LLM Generated Captions indicate whether the dataset contains captions generated by LLMs such as ChatGPT.} 
\centering
\label{tab:datasets}
{\setlength{\tabcolsep}{2.5pt}
\scalebox{0.8}{
\begin{tabular}{@{}llllllllll@{}}
\toprule
\multicolumn{1}{P{65}{2.2cm}}{Name Dataset} & \multicolumn{1}{P{65}{2.2cm}}{Use Case} & \multicolumn{1}{P{65}{2.2cm}}{Number of Captions} & \multicolumn{1}{P{65}{2.2cm}}{Avg Number of Characters with Std Dev} & \multicolumn{1}{P{65}{2.2cm}}{Unique Words} & \multicolumn{1}{P{65}{2.2cm}}{Number of Sounds} & \multicolumn{1}{P{65}{2cm}}{Total Hours of All Audio} & \multicolumn{1}{P{65}{2.2cm}}{Dataset Size (GB)} & \multicolumn{1}{P{65}{2.2cm}}{Easily Accessible} & \multicolumn{1}{P{65}{2.2cm}}{LLM Generated Captions} \\ \midrule
\multicolumn{10}{c}{\textbf{Youtube-based Datasets}} \\ \hline
\textbf{AudioSet \cite{gemmekeAudioSetOntology2017}}& Sound vocabulary based on an ontology & 2,084,317& 19.97 $\pm$ 18.92& 1,067  & 1,955,135  & 5,383.11 & 1,155.23  & \NOK   & \NOK   \\
\hspace{3mm} \textbf{AudioSet Strong} \cite{hersheyBenefitTemporallyStrongLabels2021}  & Improved temporal precision of labels  & 1,074,359& 14.7 $\pm$ 8.88  & 645 & 120,080 & 329.23   & 44.78  & \NOK   & \NOK   \\
\hspace{3mm} \textbf{AF-AudioSet} \cite{kongImprovingTextAudioModels2024} & Large scale with synthetic captions   & 696,079  & 143.65 $\pm$ 57.58 & 18,498 & 331,421 & 912.51   & 195.83 & \NOK   & \OK  \\
\hspace{3mm} \textbf{ACalt4} \cite{niizumiM2DCLAPMaskedModeling2024}  & Provides four captions per audio  & 167,140  & 63.7 $\pm$ 29.89 & 12,338 & 30,116  & 82.51& 17.71  & \NOK   & \OK  \\
\hspace{3mm} \textbf{EzAudioCaps} \cite{haiEzAudioEnhancingTextaudio2024} & Diverse synthetic captions  & 1,088,808& 81.43 $\pm$ 38.8 & 28,976 & 849,816 & 2,344.60 & 472.99 & \NOK   & \OK  \\
\hspace{3mm} \textbf{AudioSet Dialogues} \cite{goelAudioDialoguesDialogues2024a}& Dialogues for audio understanding  & 172,566  & &  & 78,084  &   &  & \NOK   & \OK  \\
\hspace{3mm} \textbf{AudioTime} \cite{xieAudioTimeTemporallyalignedAudiotext2024}   & Temporally aligned annotations  & 22,000   & 77.73 $\pm$ 27.62& 13,584 & 22,000  & 59.18& 12.70& \OK & \OK  \\
\hspace{3mm} \textbf{AudioSetMix} \cite{xuAudioSetMixEnhancingAudioLanguage2024a}   & Audio-language learning with modifiers & 49,971   & 20.00&  &   & 49,971   & 138.81 & \NOK   & \OK  \\
\hspace{3mm} \textbf{Sound-VECaps$_A$} \cite{yuanSoundVECapsImprovingAudio2024} & Improved performance on complex prompts   & 1,657,029& 180.72 $\pm$ 46.87 & 141,558& 1,645,920  & 4,531.74 & 972.52 & \NOK   & \OK  \\
\textbf{DAQA} \cite{fayekTemporalReasoningAudio2020}& Synthetic dataset for temporal reasoning  & 598,878  & 77.24 $\pm$ 19.72& 271 & 100,000 & 2,089.36 & 224.17 & \NOK   & \NOK \\
\textbf{CompA-R} \cite{ghoshGAMALargeAudioLanguage2024a}  & Synthetic dataset for complex reasoning   & 200,205  & 284.32 $\pm$ 58.05 & 49,039 & 62,617  & 171.98   & 36.91  & \OK & \OK \\
\textbf{MULTIS} \cite{zhaoChatBridgeBridgingModalities2023} & Dialogue dataset for instruction tuning   & 8,835 & 116.18 $\pm$ 48.01 & 5,156  & 7,645& 21.07& 4.52& \OK & \OK  \\
\textbf{SpatialSoundQA} \cite{zhengBATLearningReason2024} & Spatial reasoning for 3D sound& 1,749,114& 100.17 $\pm$ 30.25 & 1,639  & 1,749,114  & 4,858.65 & 2,085.18  & \NOK   & \OK  \\
\textbf{AudioCaps} \cite{kimAudioCapsGeneratingCaptions2019}& First large-scale captioning& 57,188   & 49.22 $\pm$ 24.09& 6,623  & 36,706  & 100.55   & 21.58  & \NOK   & \NOK   \\
\hspace{3mm} \textbf{Audio-ALPACA} \cite{majumderTango2Aligning2024}  & Winner and loser audios  & 15,025   & 56.94 $\pm$ 24.96& 2,795  & 15,025  & 42.75& 4.59& \OK & \OK  \\
\hspace{3mm} \textbf{AudioHallucination} \cite{kuanUnderstandingSoundsMissing2024}  & Object hallucination in ALMs & 92,643   & 51.4 $\pm$ 5.71  & 1,113  & 686  & 1.88 & 0.40 & \NOK   & \NOK   \\
\hspace{3mm} \textbf{BATON} \cite{liaoBATONAligningTexttoaudio2024}  & Human preference feedback for TTA & 4,509 & 67.59 $\pm$ 15.36& 1,083  & 5,990& 17.04& 1.69& \OK & \OK  \\
\hspace{3mm} \textbf{TextToAudioGrounding} \cite{xuTextAudioGroundingBuilding2021}   & Fine-grained temporal labels & 4,750 & 52.27 $\pm$ 24.33& 1,715  & 4,588& 12.71& 2.73& \OK & \NOK   \\
\hspace{3mm} \textbf{AQUALLM} \cite{beheraAQUALLMAudioQuestion2023a} & Large, high-quality, diverse AQA datasets & 1,435,785 & & & 61,167 & & & \NOK & \OK \\
\textbf{VGGSound} \cite{chenVGGSoundLargescaleAudioVisual2020}  & Large, audio-visual correspondence focus  & 199,467  & 16.52 $\pm$ 5.49 & 475 & 197,956 & 548.63   & 75.38  & \OK & \NOK \\
\hspace{3mm} \textbf{VGGSound-Caption} \cite{yangDrawAudioLeveraging2024} &  Language prompts for Video to Audio & 178,730  & &  & 178,730 & 496.47   &  & \NOK   & \OK  \\
\textbf{Auto-ACD} \cite{sunAutoACDLargescaleDataset2024} & Diverse with environmental context & 1,922,878& 103.97 $\pm$ 17.33 & 43,802 & 1,922,878  & 5,245.78 & 1,239.19  & \NOK   & \OK  \\
\textbf{FAVDBench Audio} \cite{shenFinegrainedAudibleVideo2023} & Diverse life-related scenes  & 9,643 & 78.56 $\pm$ 46.75& 6,433  & 10,000  & 21.55& 13.87  & \OK & \NOK   \\ \hline
\multicolumn{10}{c}{\textbf{Freesound-based Datasets}} \\ \hline
\textbf{Freesound} \cite{fontcorberaFreesoundTechnicalDemo2013} & Large, diverse, and expandable  & 1,018,339& 43.81 $\pm$ 43.17& 372,319& 515,580 & 3,003.35 & 645.89 & \OK & \NOK   \\
\textbf{FSD50k Title+Desc} \cite{fonsecaFSD50KOpenDataset2022}  & Open, large vocabulary, expandable & 51,197   & 224.22 $\pm$ 277.64    & 86,349 & 51,197  & 108.59   & 32.12  & \OK & \NOK   \\
\textbf{AudioDiffCaps} \cite{takeuchiAudioDifferenceCaptioning2023}& Differences between audios  & 37,488   & 41.11 $\pm$ 18.16& 973 & 7,756& 21.43& 12.68  & \NOK   & \NOK \\
\textbf{Clotho} \cite{drossosClothoAudioCaptioning2020}   & Focus on diversity and caption ambiguity & 29,645   & 62.84 $\pm$ 15.72& 8,168  & 6,972& 43.64& 12.91  & \OK & \NOK   \\
\hspace{3mm} \textbf{Revised-Clotho} \cite{xiaoEnsembleSystemsContrastive2023}  & Revision of inaccurate captions & 29,645   & & 8,168  & 6,972& 43.64& 12.91  & \NOK   & \OK  \\
\hspace{3mm} \textbf{Clotho-Detail} \cite{zhaoBuboGPTEnablingVisual2023}  & Expressive and descriptive captions   & 3,939 & 324.83 $\pm$ 48.08 & 4,407  & 3,939& 24.55& 7.26& \OK & \OK  \\
\hspace{3mm} \textbf{ClothoV2\_GPT} \cite{primusAdvancingNaturalLanguageBased2023}& Augments ClothoV2 with keywords.  & 95,981   & 52.61 $\pm$ 14.86& 23,676 & 3,839& 23.99& 7.09& \OK & \OK  \\
\hspace{3mm} \textbf{Clotho-ChatGPT-Mixup} \cite{wuImprovingAudioCaptioning2024}  & Increased complexity and diversity.& 50,000   & 100.39 $\pm$ 20.08 & 10,308 & 50,000  & 328.82   & 35.28  & \OK & \OK  \\
\hspace{3mm} \textbf{Clotho-Moment} \cite{munakataLanguagebasedAudioMoment2024} & Retrieves specific moments from long audio & 44,261 & 13.33 $\pm$ 2.79 & 7,808 & 49,842 & 178.25 & 830.7 & \OK & \NOK \\
\hspace{3mm} \textbf{Clotho Entailment} \cite{deshmukhAudioEntailmentAssessing2024}  & Evaluate deductive reasoning & 17,787   & 82.48 $\pm$ 19.58& 7,745  & 5,929& 37.06& 10.96  & \OK & \OK  \\
\textbf{ClothoAQA} \cite{lippingClothoAQACrowdsourcedDataset2022}  & Crowdsourced question answering & 35,838   & 35.40 $\pm$ 11.97 & 4,407  & 1,991& 12.46& 3.68& \OK & \NOK   \\
\hspace{3mm} \textbf{mClothoAQA} \cite{beheraMultiLingualAudioQuestion2023}  & Multilingual audio question answering & 203,082  & 33.63 $\pm$ 11.49& 27,522 & 1,991& 12.46& 3.68& \OK & \NOK \\
\hspace{3mm} \textbf{NLAR} \cite{liangAcousticPromptTuning2023} & Natural language audio reasoning & 200 & & & 200 & 1.2 & & \NOK & \OK \\
\textbf{WavCaps} \cite{meiWavCapsChatGPTAssistedWeaklyLabelled2023}& Weakly-labeled GPT-generated captioning   & 403,050  & 44.87 $\pm$ 21.18& 43,802 & 403,050 & 1,119.58 & 264.41 & \OK & \OK  \\
\hspace{3mm} \textbf{AudioCondition} \cite{guoAudioGenerationMultiple2024}& With text and control conditions   & 92,065   & 351.02 $\pm$ 275.91    & 43,802 & 92,065  & 255.74   & 60.37  & \NOK   & \OK  \\
\textbf{ESC50} \cite{piczakESCDatasetEnvironmental2015} & With human accuracy estimates & 2,000 & 9.04 $\pm$ 3.88 & 50 & 2,000 & 2.78 & 0.82 & \OK & \NOK \\
\hspace{3mm} \textbf{SynCaps} \cite{oncescuDissectingTemporalUnderstanding2024a}& Synthetic, focused on temporal cues& 5,370 & 39.55 $\pm$ 5.16 & 101 & 5,370& 14.92& 4.41& \OK & \OK  \\
\textbf{PicoAudio} \cite{xiePicoaudioEnablingPrecise2024} & Precise temporal control of audio events  & 12,400   & 40.42 $\pm$ 21.46& 12,414 & 5,600& 13.93& 2.99& \OK & \OK  \\
\textbf{RDATS} \cite{xuDetailedAudiotextData2024} & Controls and describes detailed audio aspects & 1,966 & 13.66 $\pm$ 4.38 & 834 & 2,000 & 6.86 & 1.47 & \NOK & \OK \\
\textbf{SoundSCaper} \cite{houSoundscapeCaptioningUsing2024a}& Acoustic, event, and affective information& 25,440 & & & 25,440 & 212 & 125.38 & \NOK & \OK \\
\textbf{DCASE LASS} \cite{liuSeparateWhatYou2022, liuSeparateAnythingYou2023a} & Dataset for DCASE Task on LASS  & 60,397   & 68.84 $\pm$ 22.71& 8,204  & 58,397  & 128.59   & 34.26  & \OK & \OK  \\ \hline
\multicolumn{10}{c}{\textbf{Sound Effect Datasets}} \\ \hline
\textbf{LAION-Audio-630k} \cite{wuLargeScaleContrastiveLanguageAudio2023}   & Large dataset for CLAP training  & 1,213,235& 45.56 $\pm$ 43.71& 391,541& 391,541 & 4,017& 867.07 & \NOK   & \OK  \\
\hspace{3mm} \textbf{Audiostock} \cite{AudiostockHighQualityRoyaltyFree}  & Royalty free sound effects   & 10,001   & 35.36 $\pm$ 14.59& 6,710  & 10,001  & 20.89& 5.79& \OK & \NOK   \\
\hspace{3mm} \textbf{Epidemic Sound} \cite{BringYourStory} & Large catalog of sound effects & 151,290  & 56.33 $\pm$ 47.83& 16,750 & 75,645  & 220.41   & 60.10& \OK & \OK  \\
\hspace{3mm} \textbf{Free To Use Sounds} \cite{FreeSoundClips} & Sound effects from field recordings& 6,370 & 93.18 $\pm$ 45.33& 4,979  & 6,370& 175.73   & 50.69  & \OK & \NOK   \\
\hspace{3mm} \textbf{ODEON (Paramount)} \cite{ODEONCinematicSound} & Cinematic sound effects   & 5,212 & 10.98 $\pm$ 4.38 & 3,709  & 5,212& 20.01& 2.03& \OK & \NOK   \\
\hspace{3mm} \textbf{Sonniss Game Effects} \cite{FreeSoundClips} & Sound effects for Game Audio & 5,561 & 42.37 $\pm$ 22.92& 7,854  & 5,561& 84.61& 22.50& \OK & \NOK   \\
\hspace{3mm} \textbf{We Sound Effects} \cite{HomeWeSound2023} & Defunct sound effect library   & 5,561 & 41.40 $\pm$ 21.58 & 7,854  & 5,561& 84.61& 22.50& \OK & \NOK   \\
\textbf{WavText5K} \cite{deshmukhAudioRetrievalWavtext5k2022}  & Improves audio-retrieval performance   & 4,348 & 73.00 $\pm$ 49.01   & 8,096  & 4,525& 25.48& 14.43  & \OK & \NOK   \\
\hspace{3mm} \textbf{BigSoundBank} \cite{sardinFreeSoundEffects} & Including ambiences and soundscapes & 2290 & 51.59 $\pm$ 34.38 & 3459 & 2467 & 19.34 & 11.03 & \NOK & \NOK \\
\hspace{3mm} \textbf{SoundBible} \cite{FreeSoundClips}& Free sound effects for multimedia projects& 1,232 & 31.95 $\pm$ 12.28& 1,827  & 1,320& 4.74 & 0.55& \OK & \OK  \\
\textbf{BBC Sound Effects \cite{BBCSoundEffects}}  & Sound effects from  TV and Radio   & 15,968   & 56.44 $\pm$ 27.72& 13,278 & 15,973  & 479.99   & 77.25  & \OK & \NOK   \\
\hspace{3mm} \textbf{SoundDescs} \cite{koepkeAudioRetrievalNatural2022}   & Large vocabulary, varied durations & 32,979   & 104.75 $\pm$ 52.23 & 26,362 & 32,836  & 1,056.63 & 621.77 & \NOK   & \NOK   \\
\textbf{Adobe Audition SFX} \cite{adobecreativecloudAdobeAuditionSound} & Uncompressed royalty-free sound effects   & 9,524 & 45.07 $\pm$ 13.19& 2,871  & 9,528& 22.42& 14.62  & \OK & \NOK   \\
\textbf{Sound Jay} \cite{FreeSoundEffects}& Free and royalty-free sound effects& 2,239 & 32.67 $\pm$ 13.28& 1,028  & 1,139& 3.57 & 2.72& \OK & \NOK   \\
\textbf{Zapsplat} \cite{DownloadFREESound} & Professionally recorded sound effects& 64,421   & 8.51 $\pm$ 2.97  & 11,040 & 64,421  & 117.67   & 27.78  & \NOK   & \NOK   \\
\textbf{Pro Sound Effects} \cite{priceProSoundEffects} & Large commercial sound effect library & 430,434 & 74.67 $\pm$ 43.77 & 51,182 & 430,434 & 4,470.56 & 5,964.95 & \NOK & \NOK \\ \hline
\multicolumn{10}{c}{\textbf{Field Recording Datasets}} \\ \hline
\textbf{MACS \cite{martin-moratoWhatGroundTruth2021, martinmoratoDiversityBiasAudio2021}} & Metric-based reliable ground truth & 17,275   & 54.26 $\pm$ 21.65& 2,775  & 14,400  & 10.92& 39.56  & \OK & \NOK   \\
\textbf{AudioLog} \cite{baiAudioLogLLMsPoweredLong2023} & Temporal information of long audio & 7,909 & & & 7,909 & 33.66 & 29.87 & \NOK & \OK \\
\textbf{AnimalSpeak} \cite{robinsonTransferableModelsBioacoustics2024} & Large bio-acoustic dataset& 895,190  & 39.38 $\pm$ 34.81& 42,968 & 878,359 & 11,841.12   & 1,131.02  & \NOK   & \OK  \\
\textbf{SoundingEarth} \cite{heidlerSelfsupervisedAudiovisualRepresentation2023} & Urban sounds for remote sensing & 51,058   & 220.96 $\pm$ 195.34& 156,983& 51,078  & 3,453.36 & 400.69 & \NOK   & \NOK   \\
\textbf{AudioEgoVLP} \cite{oncescuSOUNDAPPROACHUsing2024a}   & Audio from egocentric viewpoints & 8,954 & 66.76 $\pm$ 23.63& 4.23   & 3,346& 12.35& 3.65& \OK & \OK  \\
\textbf{MIMII-Change} \cite{tsubakiAudiochangeCaptioningExplain2023} &  Changes between two audio samples & 17,999 & & & 17,999 & 50 & 5.36 & \NOK & \NOK \\
\textbf{CAPTDURE} \cite{okamotoCAPTDURECaptionedSound2023}   & Single-source sound captions  & 8,034 & 72.96 $\pm$ 39.70 & 6,195  & 2,088& 4.90  & 2.67& \OK & \NOK   \\ \hline
\multicolumn{10}{c}{\textbf{Movies \& TV Show Datasets}} \\ \hline
\textbf{AudioCaption} \cite{wuAudioCaptionListen2019, xuAudioCaptionCar2021} & Specifically for Mandarin Chinese & 29,117 & 92.10 $\pm$ 41.89 & 8,542 & 7,336 & 21.43 & 12.96 & \OK & \NOK \\
\textbf{SAM-S} \cite{hebbarDatasetAudioVisualSound2023} & Human-centric and movie-specific sounds & 116,000 & & & 110,000 & & & \NOK & \OK \\ \hline
\end{tabular}
}
}
\end{table*}

\subsection{YouTube-based datasets} \label{sec:youtube}
YouTube-based datasets, such as AudioSet and VGGSound, have been instrumental in the development of audio captioning models. However, using these datasets comes with specific challenges due to YouTube's policies. Content creators can license their videos under the Creative Commons CC BY license, allowing reuse if they have the rights to the material. Despite this, YouTube's Terms of Service restrict content download unless explicitly authorized or permitted by YouTube.

As a result, researchers cannot directly share audio files. Instead, YouTube-derived datasets typically include only video links with the corresponding start and end timestamps. Researchers also provide scripts for scraping these videos. Consequently, anyone looking to replicate the models must scrape the data themselves.

A further complication arises from \textit{link rot}, where video owners may delete their videos, or YouTube may remove them for policy violations. This leads to the unavailability of some portions of the dataset, which complicates efforts to reproduce research results or train new models.

\subsubsection{AudioSet}
Launched by Google in 2017 \cite{gemmekeAudioSetOntology2017}, AudioSet is one of the first large-scale datasets for audio classification, containing 2,085,544 audio clips, each approximately 10 seconds in duration. AudioSet features 527 categories of audio events, each manually annotated using Amazon Mechanical Turk, providing multi-label annotations. For creating audio-language models, various researchers concatenate the audio event labels together to form a caption.

The categories are organized into an ontology with top-level labels such as Human sounds, Source-ambiguous sounds, Animals, Sounds of things, Music, Natural sounds, and Environmental and background sounds. AudioSet is divided into balanced and unbalanced training sets, with the balanced set comprising about 1\% of the total and designed to ensure at least 59 examples per category. The unbalanced set contains the majority of the examples, particularly dominated by categories such as music and speech.

Due to the transient nature of YouTube content, the exact composition of AudioSet varies over time, as videos are sometimes removed from the platform. As of March 2024, about 18\% of AudioSet samples are no longer available \cite{tsubakiRefiningKnowledgeTransfer2024}. This variation can affect the size of the dataset used in different studies by up to $ \pm 5\%$, potentially influencing outcomes and reported metrics such as mean average precision (mAP) \cite{liRobustLargescaleAudio2023}.

Several other derivative datasets have been created from AudioSet. These datasets include the following:

\textbf{AudioSet Strong} \cite{hersheyBenefitTemporallyStrongLabels2021}, which is a direct derivative, provides enhanced annotations with \textit{strong} labels that indicate the exact start and end times of each sound event within the clips. This allows for more precise training on sound event detection tasks. However, it covers only 356 of the original 527 AudioSet classes. Also, these datasets are not necessarily effective for audio-language learning since the dataset only provides specific sound events per start-end time, instead of full sentences. 

\textbf{AF-AudioSet} \cite{kongImprovingTextAudioModels2024} is a synthetic audio captioning dataset based on AudioSet produced with a data labeling pipeline. The pipeline includes the audio-language model Audio Flamingo \cite{kongAudioFlamingoNovel2024} that is prompted with the audio clips and the instruction "Can you briefly describe what you hear in this audio?" to generate 20 captions per audio with sampling using $\text{top-}k = 50$ and $\text{top-}p = 95\%$. Afterwards, CLAP \cite{wuLargeScaleContrastiveLanguageAudio2023} is used between the audio and the caption to rank and filter the caption, and the top three most correlated captions are selected, removing captions having similarities below $35\%$. Although this approach provides multiple high-quality synthetic captions per audio clip, their synthetic nature implies they could potentially contain biases from the audio-language model.

\textbf{AudioCaps Alternative 4 Captions (ACalt4)} \cite{niizumiM2DCLAPMaskedModeling2024} is a dataset that uses LLMs to generate captions in a similar way to Auto-ACD (see \ref{sec:autoacd}), where a frame is taken from the video of the AudioSet and VGGSound datasets, and an image captioning algorithm generates the image caption \cite{sunAutoACDLargescaleDataset2024}. A ChatGPT LLM then generates a new caption given the image caption and AudioSet labels. While this approach generates four different captions per audio using both visual and audio information, the dataset's reliance on video frames for caption generation may introduce unrelevant visual biases.

\textbf{EzAudioCaps} originates from four datasets: AudioSet, AudioSet Strong, VGGSound and AudioCaps, part of the dataset is based on Auto-ACD. A large part of EzAudioCaps includes two synthetically generated datasets AS-Qwen-Caps and AS-SL-GPT4-Caps based on AudioSet \cite{haiEzAudioEnhancingTextaudio2024}. AS-Qwen-Caps is a dataset with captions generated using Qwen-Audio \cite{chuQwenAudioAdvancingUniversal2023}, one of the leading audio-language models. AS-SL-GPT4-Caps is a dataset generated using GPT-4o-mini, based on the temporally-resolved annotations from AudioSet Strong. It is a large-scale, albeit noisy, dataset that combines multiple other larger-scale datasets.

\textbf{AudioSet Dialogues} \cite{goelAudioDialoguesDialogues2024a} is a dialogue-based dataset of multiturn sentences and comparison question-answer pairs for general sounds, although with low granularity. It is generated using a prompting-based approach with GPT-4 and its source data was from AudioSet. The dataset contains samples with one to four dialogues covering general sounds and focuses on creating strong correlations between rounds of dialogue through the presence of pronouns, follow-up questions, and complex contexts. The paper also describes a dataset that covers the music domain.

\textbf{AudioTime} \cite{xieAudioTimeTemporallyalignedAudiotext2024} is a dataset for the generation of temporal-controlled text audio. AudioTime includes timestamps, duration, frequency, and ordering information to help control audio generation. The dataset filters AudioSet Strong clips using CLAP similarity scores greater than 0.3 for segmented sound events. The TextToAudioGrounding model \cite{xuTextAudioGroundingBuilding2021} removes segments with scores below 0.6 when using event labels as queries.

\textbf{AudioSetMix} \cite{xuAudioSetMixEnhancingAudioLanguage2024a} is an audio capture dataset generated by applying audio transformations (speed, pitch, volume, duration, mixing and concatenation) to several clips together from the AudioSet Strong dataset, combined with an LLM to form a weakly annotated caption. In particular, the dataset puts emphasis on the ordering, and the LLM is prompted to put the temporal information of the audio transformations into the caption.

\textbf{Sound-VECaps} \cite{yuanSoundVECapsImprovingAudio2024}, and its derivative \textbf{AudioCaps-Enhanced}, are AudioSet-based datasets that provide rich captions with audio event orders, locations, and environmental information to improve audio generation models. The captions are generated using a visual captioning model named CogVLM \cite{wangCogVLMVisualExpert2024} that processes each second of video, although this visual-based approach increases data complexity during inference. Llama3 \cite{llama3teamLlama3Herd2024} is prompted together with an audio caption using EnCLAP \cite{kimEnCLAPCombiningNeural2024} and its AudioSet label. The Sound-VECaps provides two different categories of datasets. Sound-VECaps$_F$ describes full detailed information, including visual features, for example, texts, names and colors. While Sound-VECaps$_A$ removes visual-only information and contains only audible contents or environmental-descriptive information.

\textbf{DAQA}\label{sec:daqa}: The Diagnostic Audio Question Answering (DAQA) dataset \cite{fayekTemporalReasoningAudio2020} by Meta (Facebook), provides a controlled environment to evaluate temporal reasoning capabilities of machine learning models engaged in Audio Question Answering (AQA). This dataset comprises sequences synthesized from 400 distinct natural sound events, with programmatically generated questions and answers that minimize biases. The audio content is sourced primarily from the AudioSet database, along with recordings made specifically for this dataset. However, due to its synthetic nature, DAQA may not fully capture the complexity of real-world audio scenarios. Moreover, it requires a large amount of computational resources to generate the dataset.

\textbf{CompA-R}: CompA-R is a large and diverse instruction-tuning dataset synthesized by GPT4 to enable complex reasoning in audio-language models on audio-related metadata \cite{ghoshGAMALargeAudioLanguage2024a}. Each sample in CompA-R includes multi-aspect information on the audio, such as event tags and temporal markers. Its testing dataset, CompA-R-test, is a human-annotated evaluation set built to assess the reasoning capabilities of audio-language models on audio question-answering tasks. This dataset serves as a benchmark to test models fine-tuned on CompA-R for their ability to infer detailed scene descriptions based on complex audio input. The dataset is based on AudioSet Strong.

\textbf{MULTIS} \citet{zhaoChatBridgeBridgingModalities2023} introduced the Audio Conversation 10k as part of the MULTIS datasets, which focuses on multimodal instructional tasks. This particular dataset is formatted for instruction tuning and includes audio-language pairs aimed at enhancing the capabilities of the large language model in audio-question-answering scenarios. An example interaction in this dataset features a human asking "What is making the sound in the audio?" to which the model responds, "The sound in the audio is made by a waterfall pouring into a stream." All audio clips in this dataset are sourced from AudioSet.

\textbf{SpatialSoundQA} \cite{zhengBATLearningReason2024} developed a spatial sound question-answering dataset for 3D audio understanding using natural language. The dataset provides a comprehensive platform for evaluating spatial audio perception and reasoning tasks such as sound event detection and direction of arrival estimation. The data is originally from AudioSet and adapted to spatial audio using the SoundSpaces 2.0 simulator \cite{chenSoundSpaces20Simulation2022}. It is mainly used for spatial question answering, and models trained on this dataset may not generalize well to non-spatial audio understanding tasks.

\subsubsection{AudioCaps} \label{sec:audiocaps}
AudioCaps is a research dataset in audio-language learning sourced from AudioSet \cite{kimAudioCapsGeneratingCaptions2019}. Each audio clip is 10 seconds long. The dataset is widely used as a training set and as a benchmark in audio captioning research, though its captions tend to be simple and may not capture all audio nuances. The dataset is divided into training, validation, and test sets. Due to its YouTube source, the size of the dataset has decreased over time. The validation and test sets provide five captions per audio clip. The training set includes only one caption per audio. The captions of AudioCaps are created by workers on the Amazon Mechanical Turk platform.

A notable issue within AudioCaps is the presence of duplicate samples. Using clustering techniques, the researchers identified 257 audio clips with repetitive elements. These duplicates include identical introductory audio effects. This redundancy could affect the training and performance evaluations of the models developed using the dataset \cite{braliosGenerationReplicationAuscultating2023}.

Several other derivative datasets have been created from AudioCaps. These datasets include the following:

\textbf{Audio-ALPACA} is a preference dataset based on AudioCaps used in TANGO 2 \cite{majumderTango2Aligning2024} to enhance audio generation. Provides preference data derived from text prompts and their corresponding audio samples, although the samples are not always strongly aligned to their text prompts. The dataset creation involves generating audio samples from text prompts using a pre-trained Tango model. The dataset employs three distinct strategies to ensure diversity and complexity. The first strategy generates multiple audio samples with varying denoising steps from the AudioCaps captions. The second strategy introduces semantic changes to the original captions using GPT-4. This strategy creates audio from both the original and altered captions to capture nuances in audio-text alignment. The third strategy modifies the sequence of events in the captions to test the temporal relationships in audio generation. 

\textbf{AudioHallucination} \cite{kuanUnderstandingSoundsMissing2024} adapted AudioCaps to a question-answering task, providing a structured approach to assess object hallucination in audio-language models. It extracted objects from the captions using NLP techniques and formatted them into five prompts to benchmark audio-language models. The datasets are separated into 3 categories, based on the type of sampling strategy used to create the questions: random sampling, popular sampling and adversarial sampling. Random sampling samples k objects that are not present in the current audio. Popular sampling selects the top-k most frequent objects. Adversarial sampling ranks all objects by their frequency of co-occurrence.

\textbf{BATON} \cite{liaoBATONAligningTexttoaudio2024} is a relatively small dataset created to train an audio diffusion model. Although its size could limit generalizability, the dataset incorporates human preference feedback to align text-to-audio models with nuanced human judgments. The dataset construction began by selecting the top-200 most frequent categories from AudioCaps and combining them in pairs and triplets. GPT-4 then generated sentences connecting these category combinations using conjunctions like "and" and "followed by". Expert human annotators manually rephrased a subset of these generated sentences. In the final dataset, each caption is used to generate 5 different audio samples using the TANGO audio generation model \cite{ghosalTextaudioGenerationUsing2023}. The dataset is evenly split between human-annotated and LLM-generated samples.

\textbf{TextToAudioGrounding}\cite{xuTextAudioGroundingBuilding2021} is a dataset specifically designed for Text-To-Audio Grounding, a task that aims to predict the onsets and offsets of sound events described by natural language (see Section \ref{sec:lass}). The dataset features caption-oriented sound event tagging for each audio file and provides segmentation timestamps for each identified sound event. It was created by initially extracting sound event phrases from AudioSet and subsequently refining these through manual phrase merging.

\textbf{AQUALLM} \cite{beheraAQUALLMAudioQuestion2023a}: The AQUALLM dataset comprises three subsets: AQUALLM-AudioCaps, AQUALLM-Clotho, and AQUALLM-MACS. These subsets were developed using an automated framework designed to ensure breadth, quality and diversity in the content provided. The framework consists of four modules. The first module generates an initial set of potential responses from the captions following predefined rules. The second module uses an LLM to formulate a question for each potential answer. The third module validates each question-answer pair using an LLM. Successful pairs are then coupled with the corresponding audio clip. The final step involves the Question Paraphrasing Module (QPM), which paraphrases the generated questions.

\subsubsection{VGGSound}
VGGSound is a large-scale automatically generated dataset \cite{chenVGGSoundLargescaleAudioVisual2020}. The dataset was created using a combination of computer vision and machine listening techniques to automatically identify and verify audio-visual correspondences in videos. The audio events cover a wide range of categories including human actions, animal sounds, musical instruments, vehicles, and natural phenomena.

\textbf{VGGSound-Caption}: VGGSound has been adapted in Audio-Language Learning for training an audio generation model with video and loudness map as input \cite{yangDrawAudioLeveraging2024}.

\subsubsection{Auto-ACD} \label{sec:autoacd}
The Auto-ACD dataset \cite{sunAutoACDLargescaleDataset2024} is a large-scale collection based on VGGSound \cite{chenVGGSoundLargescaleAudioVisual2020} and AudioSet \cite{gemmekeAudioSetOntology2017} that provides detailed captions, although it contains many noisy entries requiring filtering. Also, since the dataset relies on visual features, it may contain biases from the visual domain. This dataset improves the original keywords from these sources using rich audiovisual information processed through six different models. Specifically, four algorithms analyze the middle frame of the video as an image: (1) BLIP-2 for image captioning, (2) Grounding Dino for object detection, (3) OpenAI's CLIP model, which categorizes images into 1000 ImageNet categories using the prompt "a photo of a {label}", and (4) PlaceCNN for place recognition.

\subsubsection{FAVDBench Audio}
FAVDBench dataset is a specialized dataset to support the fine-grained audible video description task (FAVD) \cite{shenFinegrainedAudibleVideo2023}. This dataset includes video clips, each accompanied by a summary, 4-6 visual descriptions, and 1-2 audio-related descriptions. Although visual descriptions are primarily tailored for visual understanding and are not ideal for Audio-Language Learning, audio-related descriptions are specifically crafted to capture sounds and auditory events that are discernible without visual context. For this reason, audio-related descriptions are used in this survey, while visual descriptions are not.

These videos were sourced from YouTube and are available exclusively under the Creative Commons license. The dataset was manually annotated in Chinese by a team of 60 people and subsequently translated into English. The annotation process for each video involved one primary annotator and two reviewers. In addition, each sentence in the descriptions was required to comprise more than five Chinese characters.

The same group also released another dataset, under the name TAVGBench \cite{maoTAVGBenchBenchmarkingText2024a}. This dataset focuses on audiovisual understanding and contains visual cues as well.

\subsection{Freesound-based datasets} \label{sec:freesound}
Freesound is a widely used resource in the audio domain and hosts a wide range of audio files. On Freesound, users can upload sound clips they have created or found, attaching relevant metadata such as filenames, keywords, and descriptions \citet{fontcorberaFreesoundTechnicalDemo2013}. Although user-generated metadata can vary in quality and structure, often including extraneous information or featuring audio clips of impractical lengths, the raw data typically require significant preprocessing to make it suitable for machine learning applications.

One of the main advantages of using Freesound-based datasets is the licensing. Freesound allows for the broad utilization of its sounds under various licenses, making it a valuable source for training datasets in audio-related machine learning tasks. This section will explore several Freesound-derived datasets, detailing how they have been processed and used for specific research purposes.

\subsubsection{Freesound}
Various researchers have included almost all Freesound in their training data for audio-language models \cite{wuLargeScaleContrastiveLanguageAudio2023, deshmukhPengiAudioLanguage2023, ghoshCompAAddressingGap2024}. Although Freesound continues to grow through user uploads, its user-provided descriptions are often noisy and require extensive processing. For this reason, Freesound is often used during pretraining in a self-supervised fashion.

\subsubsection{FSD50k}
FSD50k \cite{fonsecaFSD50KOpenDataset2022} is primarily a dataset for the classification of sound events that is openly distributable and contains a large vocabulary of everyday sounds. Although it does not include captions, researchers have adapted it for the training of audio-language models \cite{elizaldeClapLearningAudio2023}. This adaptation involves concatenating the title and description provided in the metadata for each audio file. Although this method generates long descriptions, it introduces a degree of noisiness due to the unstructured nature of the metadata. A limitation of the dataset is that its classes are unbalanced, and some classes have many more examples than others.

\subsubsection{AudioDiffCaps}
The AudioDiffCaps dataset \cite{takeuchiAudioDifferenceCaptioning2023} is specifically designed to describe differences between pairs of similar but distinct audio clips through human-annotated descriptions. While the dataset effectively captures subtle differences between audio pairs, its clips are artificially synthesized using the Scaper library \cite{salamonScaperLibrarySoundscape2017}, which may not fully reflect real-world conditions. The pairs are created by mixing foreground event sounds with background noises from the FSD50K and ESC-50 datasets. Each audio clip within the pairs is precisely 10 seconds in length. This dataset complements the Audio Pair with Difference dataset introduced earlier by the same researchers \cite{takeuchiIntroducingAuxiliaryText2022}.

\subsubsection{Clotho}
Clotho \cite{drossosClothoAudioCaptioning2020} is one of the most widely used datasets in the field of audio captioning. It features a training, validation and evaluation split, each ranging from 15 to 30 seconds in length. Every audio file in Clotho is paired with multiple captions created using Amazon Mechanical Turk. Additionally, Clotho includes a test set of audio samples without accompanying captions hosted on a separate Zenodo page, specifically designed for inference purposes. This dataset is also prominently used in the DCASE challenges for Automated Audio Captioning. Its advantage lies in that it contains five captions per training, validation and evaluation sample. 

A study by \citet{xieCrowdsourcingEvaluatingTextBased2023} involved a crowd-sourcing task on Mechanical Turk to evaluate the precision of 200 Clotho captions. Participants were asked to score how well the sound content matched the given caption and other relevant captions not attributed to the audio. From this study, which collected 18,204 responses from 340 workers, it was found that approximately 60\% of the audio content in Clotho accurately matches its given caption. This dataset is proposed as GrRel in subsequent work, including relevance rankings from 0 to 100 of the audio samples given a caption \cite{xieIntegratingContinuousBinary2024}.

Several researchers have developed revised or extended versions of the Clotho dataset to address specific challenges and enhance its utility for training audio captioning models. Key developments include:

\textbf{Revised-Clotho} \cite{xiaoEnsembleSystemsContrastive2023} \label{sec:revised-clotho}: This dataset is a modified version of Clotho that improves caption accuracy through GPT-3.5-based validation. While it addresses issues with captions that poorly represent audio content by evaluating and revising them based on audio file name tags, the automated revision process may introduce biases that affect the natural diversity of the original captions.

\textbf{Clotho-Detail} \cite{zhaoBuboGPTEnablingVisual2023}: Is a relatively small dataset designed for instruction tuning that provides expressive and descriptive captions for deep audio content understanding. It was generated automatically using GPT-4, which consolidated the five original captions per audio into a single comprehensive caption in a few-shot setting.

\textbf{ClothoV2\_GPT} \cite{primusAdvancingNaturalLanguageBased2023} \label{sec:clotho-gpt}: Using GPT-3.5, this version involves generating new captions for the Clotho dataset. This process used both the keywords and the original captions to produce five distinct captions for each audio file. While the additional metadata-augmented captions help reduce overfitting during fine-tuning, the GPT-generated captions only marginally improve performance since Clotho already contains diverse human-generated captions.

\textbf{Clotho-ChatGPT-Mixup} \cite{wuImprovingAudioCaptioning2024}: The winner of the DCASE Challenge 2023 on Audio Captioning uses ChatGPT to propose caption mix-ups, which increases the complexity and diversity of the training data. They provide ChatGPT two randomly sampled captions and ask it to come up with a mix of the two while keeping the number of words under 25. The audio is mixed by scaling the two waveforms to ensure that their relative root mean square energy is within $\pm 5$ dB before adding them together \cite{chenWavLMLargescaleSelfsupervised2022}.

\textbf{Clotho-Moment} \cite{munakataLanguagebasedAudioMoment2024} is a proposed dataset for Audio Moment Retrieval that enables large-scale training by simulating long recorded audio in the city. The audio and text pairs are generated by overlaying Clotho samples with the Walking Tour dataset \cite{venkataramananImageNetWorth12024} at randomly sampled intervals. While this approach allows for diverse moment annotations and robust model evaluation, the simulated nature may not perfectly capture real-world audio complexities. The resulting dataset consists of 1-minute-long samples.

\textbf{Clotho Entailment} (and AudioCaps Entailment) \cite{deshmukhAudioEntailmentAssessing2024} are two datasets to assess if a model can assess if a text description (hypothesis) of audio content can be deduced from an audio recording (premise). The dataset has three options for this. Either entailment, neutral or contradiction. For each of these three, a model should also provide its evidence. 

\subsubsection{ClothoAQA} \label{sec:clothoaqa}
ClothoAQA \cite{lippingClothoAQACrowdsourcedDataset2022} is derived from the Clotho dataset to be specifically used to answer audio questions. The dataset integrates crowd-sourced annotations through Amazon Mechanical Turk. Each audio sample is accompanied by six questions and three corresponding answers, totalling 18 question-answer pairs. Each pair of question-answers per audio was annotated by a different worker. The questions are structured to vary in difficulty: two per audio file are `yes-no' questions designed for binary responses, while the remaining four require a single-word answer, which could be a limiting factor. 

ClothoAQA subsequently has 2 datasets that are based on it, named NLAR and mClothoAQA \cite{lippingClothoAQACrowdsourcedDataset2022}:

NLAR \cite{liangAcousticPromptTuning2023} is a small dataset created for the task of natural language audio reasoning. It is based on ClothoAQA, by cleaning the dataset. NLAR uses ChatGPT-Turbo to summarize acoustic characteristics and uses ChatGPT to generate five question-answer pairs given pairs of audio and its description. 

mClothoAQA \cite{beheraMultiLingualAudioQuestion2023} expands the audio question-answering framework of ClothoAQA \cite{lippingClothoAQACrowdsourcedDataset2022} into a multilingual context. By providing audio-caption pairs in eight languages (English, French, Hindi, German, Spanish, Italian, Dutch, and Portuguese), it addresses the lack of multilingual datasets in the field. The dataset was constructed by automatically transcribing the English question-answering pairs of ClothoAQA into seven other languages using Google's machine translation API, though this automated translation process may introduce language bias issues.

\subsubsection{WavCaps} \label{sec:wavcaps}
WavCaps \cite{meiWavCapsChatGPTAssistedWeaklyLabelled2023} is a weakly labeled dataset that brings together audio clips from four diverse sources: Freesound, BBC Sound Effects, SoundBible, and AudioSet (strong labels), of which the largest portion of the sounds come from Freesound. The dataset is generated using ChatGPT from raw descriptions. Additionally, WavCaps includes both prefiltering and post-processing steps aimed at refining the dataset by eliminating undesirable data, which resulted in a high discard rate. 

\textbf{AudioCondition}: \cite{guoAudioGenerationMultiple2024} is a dataset for conditional audio generation with text and control conditions based on AudioSet Strong, with its caption from WavCaps. For each category of AudioSet Strong, a new caption, albeit limited to the category, is created with timestamps based on the information from the WavCaps caption. The dataset enables fine-grained control over audio generation by incorporating timestamp, pitch contour, and energy contour for precise and diverse audio synthesis. For audio-to-language models, this dataset is less suitable due to its unstructured text captions.

\subsubsection{ESC-50}
ESC-50 \cite{piczakESCDatasetEnvironmental2015} is a well-balanced environmental sound classification dataset that contains 2000 4-second audio clips in 50 classes of common environmental sounds. The classes are organized into five major categories: animal sounds, natural soundscapes and water sounds, human non-speech sounds, interior/domestic sounds, and exterior/urban noises. While the dataset provides high-quality labeled recordings with sound events in the foreground and limited background noise, its small size of only 40 clips per class limits its utility for complex machine learning models. The dataset was created by extracting clips from Freesound recordings and was manually annotated by multiple participants through the CrowdFlower crowdsourcing platform, achieving 81.3\% human classification accuracy. Additionally, without longer textual descriptions, the dataset's short class labels make it less suitable for audio-language models requiring more detailed captions.

\textbf{SynCaps} \cite{oncescuDissectingTemporalUnderstanding2024a} is a synthetically created and used as a fine-tuning dataset for text-to-audio and audio-to-text retrieval. An LLM generates textual descriptions in the style of AudioCaps from ESC50 sound labels. The dataset provides uniform temporal cues in the captions by concatenating two sounds with arbitrarily selected temporal order using only future and past cues. This helps models learn temporal relationships in audio, although the temporal cues are limited.

\subsubsection{PicoAudio}
Picoaudio \cite{xiePicoaudioEnablingPrecise2024} is a dataset created for evaluating the temporal control performance of audio generation models. The dataset provides high-quality, temporally aligned audio-text data that enables precise control over timestamps and frequency in audio generation tasks. Each audio has either 1, 2 or 3 sound events, with the sound events originating from Freesound. This limited number of audio events could restrict its application scope. The sound events and the on-set time are randomly assigned, with two times as many clips containing 1 or 2 sound events as 3 sound events.

\subsubsection{RDATS}
RichDetailAudioTextSimulation \cite{xuDetailedAudiotextData2024} comprises a small number of simulated audio-text pairs with rich details in four aspects: temporal relationship, loudness, speaker identity, and occurrence number. The data was generated using curated single-event sounds from Freesound, mixed and concatenated according to randomly sampled attributes, and paired with descriptions generated by ChatGPT to provide more detailed captions.

\subsubsection{SoundSCaper}
SoundScaper is a model capable of generating soundscape captions, given an audio clip \cite{houSoundscapeCaptioningUsing2024a}. It is composed of a audio classification model, capable of predicting the acoustic scene and events, together with its 8-dimensional perceived affective quality values (such as Pleasantness–Eventfulness, or Calmness–Vibrancy). A GPT3.5 LLM generates a soundscape caption based on these values. The data originate from ARAUS \cite{ooiARAUSLargescaleDataset2022}, which contains the Urban SoundScapes of the World dataset, but also Freesound and Xeno-Canto. It is used to generate soundscape captions, capturing acoustic environment information from multiple perspectives.

\subsubsection{DCASE LASS}
The Language-Queried Audio Source Separation \cite{liuSeparateWhatYou2022, liuSeparateAnythingYou2023a} is a 2024 DCASE Challenge for separating arbitrary sound sources using textual descriptions of the desired source. Its training set is based on FSD50k. The DCASE-Synth and DCASE-Real datasets are for validation and evaluation purposes. The DCASE-Real dataset consists of audio clips from FreeSound with either 2 or 3 manually annotated captions per audio. The DCASE-Synth dataset consists of synthesized mixtures, with a text description and its corresponding target source. 

\subsection{Sound effect datasets} \label{sec:sfx}
Sound effects (SFX), originally designed for multimedia enhancements, have found significant utility in the audio domain. These datasets are invaluable not just for their primary purpose - adding audio layers to visual media like comedy series or action movies - but also for their structured captions, which facilitate quick and accurate retrieval of specific sounds used when searching for specific content on SFX search websites. This structured metadata is a key advantage for training audio models as it allows for use as a caption describing the sound in these datasets.

A notable subset of SFX is the Foley sounds \cite{degotzenRealFoleySynthetic2013, amentFoleyGrailArt2014}. Foley artists typically use everyday objects to mimic the sounds associated with actions or objects in a scene. For example, the sound of breaking bones might be simulated by snapping celery sticks. This creative process not only produces similar sound effects but also provides data for training audio models to recognize and replicate these sounds. However, some Foley sounds could be sonically different sounds with the same textual representation \cite{wilkinsBridgingHighQualityAudio2023}. This section explores popular datasets that are based on these SFX sounds.

\subsubsection{LAION-Audio-630k}
LAION-Audio-630k \cite{wuLargeScaleContrastiveLanguageAudio2023} is the training set on which LAION-CLAP is trained. With 633,526 audio-text pairs, this extensive dataset with varied sources enables improved model generalization across diverse audio tasks. The following datasets are part of LAION-Audio-630k, next to Freesound and BBC SFX:

\textbf{Audiostock} \cite{AudiostockHighQualityRoyaltyFree} is a huge repository of high-quality royalty-free sound effects with a steep subscription fee of \$29.99 per month.

\textbf{Epidemic Sound} \cite{BringYourStory}, a sound effects catalog that offers royalty-free high-quality music and sound effects for videos. While it offers a subscription for \$10.99 per month for personal usage, some users have reported receiving copyright claims on platforms like YouTube despite having paid subscriptions \cite{sreyashichatterjeeEpidemicSoundReview2024}.

\textbf{Free To Use Sounds} \cite{marcelFreeUseSounds}, an indie sound effect catalog offering a \$25 all in one bundle, provides royalty-free sounds recorded worldwide. However, like other Epidemic Sound, users have reported receiving copyright strikes despite having purchased licenses \cite{lundNotUseFreetousesoundscom2023}.

\textbf{ODEON Sound Effects Library} of Paramount Motion Sound Effects \cite{ODEONCinematicSound} provides an extensive library in 53 categories, although it primarily contains movie and TV effects. The library costs \$49,95.

\textbf{Sonniss Game Effects} \cite{SONNISSProfessionalSound}: A yearly release of free game effects for the Game Developers Conference in various bundles. Although the sound effects are free, they primarily target game audio, and thus have limited diversity. The LAION-Audio-630k dataset contains the data up until 2020.

\textbf{We Sound Effects} \cite{HomeWeSound2023}: a defunct website that offered high-quality sound libraries from top sound designers. Can still be accessed through archive.org.

\subsubsection{WavText5k}
WavText5k \cite{deshmukhAudioRetrievalWavtext5k2022}, created by Microsoft, improves audio-retrieval performance on benchmarks like AudioCaps and Clotho through its contrastive learning approach. The dataset does not share audio files directly but provides links to original sounds with detailed metadata. The captions come from free-form descriptions created by the sound effect creators that can lack consistency compared to human-curated annotations. This may limit the robustness of the dataset for complex retrieval scenarios. Its audios originate from two sound effect libraries:

\textbf{BigSoundBank} \cite{sardinFreeSoundEffects}: a sound effect website by Joseph Sardin offering free, high-quality, royalty-free sounds across 82 categories and 834 subcategories. Some subcategories are in French.

\textbf{SoundBible} \cite{FreeSoundClips}: offering public domain sounds to download in both wav and mp3 format, most of them created by Daniel Simion. While each sound has a title, description and keywords, the dataset is quite small and restrictive since it primarily targets video editors, movie scorers and game designers.

\subsubsection{BBC Sound Effects} \label{sec:bbc}
BBC Sound Effects database \cite{BBCSoundEffects} offers a collection of 33,066 audio clips spread over 23 categories, such as nature, transport, and machines. This library features historical sounds from the BBC Radiophonic Workshop, the Blitz in London, various BBC TV and radio productions, and recordings from the BBC Natural History Unit archive. The accompanying captions are noisy and typically begin with nouns that identify the sound source or scene, followed by more detailed descriptions, which are often incomplete.

\textbf{SoundDescs} \label{sec:sounddescs} \cite{koepkeAudioRetrievalNatural2022} serves as a reference audio dataset. It contains a wide range of audio content and vocabulary in the descriptions. The dataset's wide variation in audio duration, with 109 files exceeding 10 minutes, helps train models to handle audio of different lengths.  The dataset is organized into training (70\%), validation and test splits (15\% each), covering 23 categories such as nature, clocks, and fire. Recent evaluations have identified data leakage in SoundDescs, including issues with duplicates, reprocessed recordings, and partial overlaps, and proposed a new split of the dataset to eliminate these issues \cite{weckDataLeakageCrossmodal2023}. As no LLM was used to generate the captions, this dataset still contains unstructured noisy captions.

\subsubsection{Adobe Audition SFX}
Adobe offers a dataset of uncompressed sound effects as part of Adobe Audition and Adobe Creative Cloud \cite{adobecreativecloudAdobeAuditionSound}. The files are grouped together by type and compressed into ZIP archives for easy download. In total, the dataset provides various different sound effects across 24 categories of audio sounds. Its filenames are used as captions, which results in noisy descriptions. The dataset has been used for training audio generation models, both latent diffusion models \cite{huangMakeAudio2TemporalEnhanced2023} and latent consistency models \cite{liuAudioLCMTextAudioGeneration2024}. However, the sound effects are only accessible to active Creative Cloud members and they can be used royalty-free in projects.

\subsubsection{Sound Jay}
Sound Jay \cite{FreeSoundEffects} is an easy to use website offering a small but diverse collection of free sound effects, though its use is limited to personal purposes. The website groups the sounds in 10 main categories, each with subcategories. It offers sounds in both WAV and MP3 format and provides captions as well as keywords for each sound.

\subsubsection{Zapsplat}
Zapsplat offers an extensive library of 150K+ free sound effects on its website under a 5\$ subscription, where a free version offers a limited 4 downloads per hour and requires attribution when using the sounds. Its sounds are divided in various categories and subcategories.

\subsubsection{Pro Sound Effects}
The Pro Sound Effects Library, a commercial resource with over a million professionally curated sounds, was the first dataset to be used in an Automated Audio Captioning setting \cite{drossosAutomatedAudioCaptioning2017}. While its complete bundle costs up to \$15,000, the library offers high-quality sounds spread over 501 categories. After an initial automatic curation process, the dataset was refined to 152,619 samples in \cite{drossosAutomatedAudioCaptioning2017}. In our analysis in \ref{tab:alm-overview}, all the website currently offers are described in Table \ref{tab:datasets}.

Subsequent studies have used this dataset for specialized applications. \citet{kilgourTextdrivenSeparationArbitrary2022} utilized 110,000 samples for the text-driven separation of sounds, a technique that improves the clarity and isolation of specific audio elements. Furthermore, \citet{miaoZeroshotTransferWildlife2023} used samples from this library to advance research in bioacoustics detection of wildlife.

\subsection{Field recording datasets}
This section explores a range of audio-language datasets that were not derived from common sources such as YouTube, Freesound, or standard sound effects websites. Instead, these datasets originate from field or studio recordings. These sources include MACS and AudioLog, both based on TUT Sound Events datasets These sources include bioacoustics websites such as AnimalSpeak, where users have recorded sounds from birds in the wild. Adaptations of existing datasets from related domains, such as MACS. Content from other video platforms, including SAM-S and AudioCaption. Self-recorded datasets, such as CAPTDURE, and collections from other online sources, such as SoundingEarth.

\subsubsection{MACS}
MACS \cite{martin-moratoWhatGroundTruth2021, martinmoratoDiversityBiasAudio2021}, a human-annotated audio dataset, features 3,930 files, each annotated with an average of 4.4 captions. Created from the TAU Urban Acoustic Scenes 2019 Dataset \cite{heittolaTAUUrbanAcoustic2019}, MACS focuses specifically on urban environments, encapsulating three distinct sound categories: airport, public square, and park sounds. Each audio file is precisely 10 seconds long, with captions averaging 9.247 words in length, contributed by 133 annotators.

\subsubsection{AudioLog}
AudioLog is a dataset for training audio-language models to generate long, description-length paragraphs on the events happening in the audio \cite{baiAudioLogLLMsPoweredLong2023}. It effectively integrates acoustic scene classification and sound event detection with contrastive learning, though its generalizability may be limited due to relying on specific acoustic scene annotations. The dataset is generated by an LLM capable of summarizing temporal information in long audio sequences. It uses the MAESTRO dataset \cite{mesarosTUTDatabaseAcoustic2016}, the TUT Acoustic Scenes 2016\&2017 and the TUT Sound Events 2016\&2017. The TUT Acoustic Scenes 2017 contains audio within two classes, "beach" and "office", that also exist in the TAU Urban Acoustic Scenes 2019 Dataset that MACS is based on.

\subsubsection{AnimalSpeak}
The AnimalSpeak dataset \cite{robinsonTransferableModelsBioacoustics2024}, is a comprehensive collection of bioacoustic sounds aimed at improving bioacoustic research. This dataset was initially compiled using unstructured labels with the assistance of ChatGPT, and further refined through screening with a language model. Captions that did not pass the screening were revised and recaptioned to ensure accuracy and relevance. By pairing over a million audio recordings with human language captions, it enables scalable analysis across diverse species, though its species distribution shows some bias toward North American and European species. Due to its many latin names, it contains novel vocabulary that may not be present in other datasets.

AnimalSpeak consists of several subsets, each sourced from distinct databases:

\begin{itemize}

\item iNaturalist: This platform offers a crowd-sourced collection of animal observations, contributing 300,267 samples to the dataset.
\item Xeno-Canto: Known for sharing wildlife sounds, particularly birds, and recently expanding to include grasshoppers and bats, this website adds 679,676 samples.
\item Watkins Marine Mammal Database: An open-access repository of marine mammal recordings collected throughout the career of William Watkins, providing 15,567 samples.
\item Animal Sound Archive: Maintained by the Museum f{\"u}r Naturkunde Berlin, this database initially contributed 21,473 samples. Unfortunately, most of this subset was lost due to a cyberattack.
\item AudioCaps: see Section \ref{sec:audiocaps}.
\end{itemize}

Each audio clip in the dataset, except those from AudioCaps, includes two captions: one referencing the animal species by its common name and the other by its scientific (Latin) name. In total, AnimalSpeak contains 1,062,983 samples, making it one of the largest bioacoustic datasets available.

\subsubsection{SoundingEarth}
The SoundingEarth dataset \cite{heidlerSelfsupervisedAudiovisualRepresentation2023} is a comprehensive collection of 50,454 geotagged ambient sound recordings from the Radio Aporee project, collectively accumulating over 3,500 hours of audio. While the dataset aims to archive field recordings globally, its reliance on Radio Aporee introduces an uneven spatial distribution that clusters heavily in Europe and the U.S., potentially limiting its global generalizability.

Each audio file in the dataset is accompanied by corresponding Google Earth images, enabling self-supervised audiovisual learning without manual annotation. This feature provides a robust foundation for tasks like remote sensing and environmental context interpretation. The recordings vary significantly in length, with a median duration of approximately 3 minutes, though some of the longest samples extend beyond 30 minutes, with noisy but manually annotated captions.

\subsubsection{AudioEpicMIR \& AudioEgoMCQ (AudioEgoVLP)}
Both the AudioEpicMIR and the AudioEgoMCQ are datasets from a proposed benchmark on audio retrieval \cite{oncescuSOUNDAPPROACHUsing2024a}. AudioEpicMIR is based on the EpicMIR \cite{damenRescalingEgocentricVision2022} task, which is based on EpicKitchens \cite{damenScalingEgocentricVision2018}. AudioEgoMCQ is based on EgoMCQ \cite{linEgocentricVideolanguagePretraining2022}, which is based on Ego4D \cite{graumanEgo4dWorld30002022}. Both provide audio descriptions from an egocentric perspective. The LLM GPT3.5 is used to generate audio captions given visual descriptions, within a prompt that also contains a few examples and an introduction to the task. Videos that did not contain any audio were removed and another LLM GPT4 is used to generate the descriptions into 3 categories: low, medium and high. These categories scale the importance of the audio for the visual task. For analysis in Table \ref{tab:datasets}, the high category is used for both datasets, to only include audio that is very informative, and combined the two datasets together.

\subsubsection{MIMII-Change} 
The MIMII-Change dataset \cite{tsubakiAudiochangeCaptioningExplain2023} is a specialized collection derived from the MIMII-DG dataset, which focuses on identifying malfunctions in industrial machines. Designed to study audio changes such as pitch and volume variations, this dataset comprises 1,500 manually annotated pairs of sounds, though it is limited to only five different types of machine sounds. For each type of sound, there are 300 pairs evenly divided between normal operating sounds and their anomalous counterparts. The unique aspect of this dataset is the use of onomatopoeia in the annotations, where the annotators describe the specific changes in sound between each pair. Consequently, each pair of sounds is accompanied by three distinct captions.

\subsubsection{CAPTDURE}
CAPTDURE \cite{okamotoCAPTDURECaptionedSound2023}, is a manually created dataset comprised of 1,044 audio recordings captured by researchers in 14 different environments, each annotated with 3-5 captions to provide a comprehensive description of the sound. It consists of single-source sounds and multiple-source sounds, each with captions and appropriateness scores for each caption. Originally crafted in Japanese and translated into English. The annotation was carried out by crowd workers who not only provided captions but also participated in a quality control process by evaluating the captions submitted by their peers.

\subsection{Movies \& TV Sound Datasets}
\subsubsection{AudioCaption}
The AudioCaption project encompasses two distinct datasets. Each dataset focuses on sounds from specific environments: one in a hospital setting and the other inside a car. This scene-specific focus helps capture sounds typical for these environments. The datasets were originally annotated in Mandarin and later translated to English using automated tools, which may introduce some translation inaccuracies. Due to their shared origins and annotation methodologies, these datasets are often used together.

\begin{itemize}
\item The Hospital dataset \cite{wuAudioCaptionListen2019}: these clips were sourced from Chinese video platforms Youku, Iqiyi and Tencent, predominantly featuring content from TV shows. The annotation process involved first asking annotators four questions about the sounds they heard, the sources of these sounds, their characteristics, and the sound's location. Responses to these queries formed the basis for the three captions assigned to each audio clip, culminating in a total of 11,121 captions. The dataset is divided into 3,337 training and 371 test clips.

\item The Car dataset \cite{xuAudioCaptionCar2021}: each clip in this dataset is annotated with five captions, making for a total of 18,010 captions. These captions were generated by five Mandarin speakers per audio clip.

\end{itemize}

\subsubsection{SAM-S}
The Subtitle-Aligned Movie Sounds (SAM-S) dataset \cite{hebbarDatasetAudioVisualSound2023} provides a scalable, semi-automatic approach to extracting sound-specific data from closed-caption transcripts of 430 Hollywood movies released between 2014 and 2018. While its closed-caption-based tagging process misses many potentially valuable audio events, the dataset still contains 116,000 captions that were originally enclosed in brackets, a method used to denote sound descriptions in film subtitles. These captions provide detailed information about the sounds, their sources, and the quality of the audio events depicted in the movies. Of the total captions, 21,000 were identified as unique. The breakdown of these captions reveals a difference in length: 1,500 consist of a single word, 11,000 contain two words, and the remaining captions extend to three words or more.

\begin{table*}[ht]
   \centering
   \caption{The sound event datasets (sorted by count) that were used in some of the major audio-language models (sorted by date). At the bottom is the total amount of sound event data these models were trained on. *Some papers only indicate the total amount of captions in their data, others the total hours of audio the data contains. In some cases both. }
   \label{tab:alm-overview}
   {\setlength{\tabcolsep}{2.5pt}
   \begin{tabular}{lllllllllllllllllllllllllllll}
   \toprule
   Dataset \hspace{5.8em} \rot{Audio-Language models} & \rot{Netease AAC \cite{yuanDCASE2021Challenge2021}} & \rot{SoundWords \cite{kilgourTextdrivenSeparationArbitrary2022}} & \rot{MS-CLAP \cite{elizaldeClapLearningAudio2023}} & \rot{AudioGen \cite{kreukAudioGenTextuallyGuided2023}} & \rot{LAION-CLAP \cite{wuLargeScaleContrastiveLanguageAudio2023}} & \rot{AudioLDM \cite{liuAudioLDMTexttoAudioGeneration2023}} & \rot{Make-An-Audio \cite{huangMakeAudioTextAudioGeneration2023}} & \rot{TangoPromptBank \cite{ghosalTextaudioGenerationUsing2023}} & \rot{ONE-PEACE \cite{wangONEPEACEExploringOne2023}} & \rot{LTU (OpenAQA) \cite{gongListenThinkUnderstand2023}} & \rot{Pengi \cite{deshmukhPengiAudioLanguage2023}} & \rot{CoDi \cite{tangAnyAnyGenerationComposable2023a}} & \rot{Make-An-Audio 2 \cite{huangMakeAudio2TemporalEnhanced2023}} & \rot{FALL-E \cite{kangFALLEFoleySound2023}} & \rot{AudioLDM 2 \cite{liuAudioLDM2Learning2024}} & \rot{Uniaudio \cite{yangUniAudioAudioFoundation2023a}} & \rot{CompA \cite{ghoshCompAAddressingGap2024}} & \rot{CLARA \cite{noriyCLARAMultilingualContrastive2023}} & \rot{SALMONN \cite{tangSALMONNGenericHearing2024}} & \rot{Auffusion \cite{xueAuffusionLeveragingPower2024}} & \rot{Audio Flamingo \cite{kongAudioFlamingoNovel2024}} & \rot{Cacophony \cite{zhuCacophonyImprovedContrastive2024}} & \rot{MINT \cite{zhaoMINTBoostingAudioLanguage2024}} & \rot{AudioLCM \cite{liuAudioLCMTextAudioGeneration2024}} & \rot{GAMA \cite{ghoshGAMALargeAudioLanguage2024a}} & \rot{AutoReCap-XL \cite{haji-aliTamingDataTransformers2024a}}& \rot{AUDIO-TPE \cite{wangAudioComposerFinegrainedAudio2024}} & \rot{Count}    \\ \midrule
   \textbf{AudioCaps} \cite{kimAudioCapsGeneratingCaptions2019}& \OK     & \OK  & \OK     & \OK & & \OK   & \OK      & \OK & \OK     & \OK & \OK   & \OK        & \OK  &  & \OK& \OK        & \OK&     & \OK    & \OK& \OK        & \OK    & & \OK & \OK    &   &  & 21    \\
   \textbf{AudioSet} \cite{gemmekeAudioSetOntology2017}& &      & & \OK        & & \OK   & \OK      & \OK & \OK     & \OK & \OK   & \OK        &      & \OK& \OK& \OK        & \OK& \OK &        &    & \OK        & \OK    & \OK     &     & \OK    & \OK      &  & 18    \\
   \textbf{Clotho} \cite{drossosClothoAudioCaptioning2020}      & &      & \OK     & \OK        & && \OK      &     & \OK     & \OK & \OK   &   & \OK  & \OK&    &   &    & \OK & \OK    &    & \OK        & \OK    & & \OK & \OK    &   &  & 14    \\
   \textbf{MACS} \cite{martin-moratoWhatGroundTruth2021, martinmoratoDiversityBiasAudio2021} & &      & \OK     &   & && \OK      &     & \OK     &     & \OK   &   & \OK  &  &    &   & \OK& \OK &        & \OK& \OK        & \OK    & & \OK & \OK    &   &  & 12    \\
   \textbf{Freesound} \cite{fontcorberaFreesoundTechnicalDemo2013}    & \OK     &      & &   & \OK   & \OK   &   &     & \OK     & \OK & \OK   & \OK        &      &  &    &   & \OK&     &        &    & \OK        & \OK    & \OK     &     & \OK    & \OK      &  & 13    \\
   \textbf{FSD50K} \cite{fonsecaFSD50KOpenDataset2022} & &      & \OK     & \OK        & && \OK      &     & & \OK & \OK   &   & \OK  &  &    &   & \OK& \OK &        &    & \OK        & & \OK     & \OK &        &   & \OK     & 12    \\
   \textbf{WavCaps} \cite{meiWavCapsChatGPTAssistedWeaklyLabelled2023}& &      & &   & &&   & \OK & &     & \OK   &   & \OK  &  & \OK& \OK        &    &     & \OK    & \OK& \OK        & & \OK     & \OK &        &   &  & 10    \\
   \textbf{BBC SFX} \cite{BBCSoundEffects}     & &      & & \OK        & \OK   & \OK   & \OK      &     & &     & \OK   & \OK        &      &  &    &   & \OK&     &        &    & \OK        & & &     & \OK    & \OK      &  & 10    \\
   \textbf{WavText5k} \cite{deshmukhAudioRetrievalWavtext5k2022}     & &      & &   & && \OK      &     & \OK     &     & \OK   &   & \OK  &  &    &   & \OK&     &        &    & \OK        & \OK    & & \OK & \OK    &   &  & 9     \\
   \textbf{Sonniss} \cite{SONNISSProfessionalSound}  & &      & & \OK        & \OK   & & \OK      &     & &     &       &   &      & \OK&    &   & \OK&     &        &    & \OK        & \OK    & &     & \OK    &   &  & 8     \\
   \textbf{VGGSound} \cite{chenVGGSoundLargescaleAudioVisual2020}     & &      & & \OK        & &&   &     & & \OK &       &   &      &  & \OK&   &    & \OK &        &    & \OK        & & \OK     &     & \OK    & \OK      &  & 8     \\
   \textbf{Audiostock} \cite{AudiostockHighQualityRoyaltyFree}  & &      & &   & \OK   & & \OK      &     & \OK     &     &       &   & \OK  &  &    &   &    &     &        &    & \OK        & \OK    & & \OK &        &   &  & 7     \\
   \textbf{Epidemic Sound} \cite{BringYourStory}     & &      & &   & \OK   & & \OK      &     & \OK     &     &       &   & \OK  &  &    &   &    &     &        &    & \OK        & \OK    & & \OK &        &   &  & 7     \\
   \textbf{ESC-50} \cite{piczakESCDatasetEnvironmental2015}     & &      & &   & && \OK      & \OK & &     &       &   & \OK  &  &    &   & \OK& \OK &        & \OK&   & & &     &        &   & \OK     & 7     \\
   \textbf{Paramount SFX} \cite{ODEONCinematicSound} & &      & & \OK        & \OK   & &   &     & &     &       &   & \OK  & \OK&    &   &    &     &        &    & \OK        & & & \OK &        &   & \OK     & 7     \\
   \textbf{UrbanSound8K} \cite{salamonDatasetTaxonomyUrban2014}& &      & &   & && \OK      & \OK & &     &       &   &      &  &    &   & \OK& \OK &        & \OK&   & & &     &        &   & \OK     & 6     \\
   \textbf{SoundBible} \cite{FreeSoundClips}& \OK     &      & &   & &&   &     & & \OK &       &   &      &  &    &   & \OK&     &        &    & \OK        & & &     & \OK    & \OK      &  & 6     \\
   \textbf{Free To Use Sounds} \cite{marcelFreeUseSounds} & &      & & \OK        & \OK   & &   &     & &     &       &   &      & \OK&    &   &    &     &        &    & \OK        & \OK    & &     &        &   &  & 5     \\
   \textbf{We Sound Effects} \cite{HomeWeSound2023}  & &      & & \OK        & \OK   & & \OK      &     & &     &       &   &      & \OK&    &   &    &     &        &    & \OK        & & &     &        &   &  & 5     \\
   \textbf{SoundDescs} \cite{koepkeAudioRetrievalNatural2022}   & &      & &   & &&   &     & \OK     &     & \OK   &   &      &  &    &   &    &     &        &    & \OK        & \OK    & &     &        &   &  & 4     \\
   \textbf{Adobe Audition SFX} \cite{adobecreativecloudAdobeAuditionSound} & &      & &   & &&   &     & &     &       &   & \OK  &  &    &   &    &     &        &    &   & & & \OK &        &   &  & 2     \\
   \textbf{Clotho-AQA} \cite{lippingClothoAQACrowdsourcedDataset2022} & &      & &   & &&   &     & &     & \OK   &   &      &  &    &   &    &     &        &    & \OK        & & &     &        &   &  & 2     \\
   \textbf{CochlScene} \cite{jeongCochlSceneAcquisitionAcoustic2022}  & &      & &   & &&   &     & &     & \OK   &   &      &  &    &   &    &     &        &    & \OK        & & &     &        &   &  & 2     \\
   \textbf{TUT} \cite{heittolaTAUUrbanAcoustic2019}  & &      & &   & &&   &     & &     &       &   & \OK  &  &    &   &    &     &        &    &   & & & \OK &        &   &  & 2     \\
   \textbf{Chime-Home} \cite{fosterChimehomeDatasetSound2015}   & &      & &   & &&   &     & &     &       &   &      &  &    &   &    &     &        &    & \OK        & & &     &        &   &  & 1     \\
   \textbf{FindSounds} \cite{FindSoundsBrowseSounds} & &      & &   & &&   &     & &     & \OK   &   &      &  &    &   &    &     &        &    &   & & &     &        &   &  & 1     \\
   \textbf{NonSpeech7K} \cite{rashidNonspeech7kDatasetClassification2023}    & &      & &   & &&   &     & &     &       &   &      &  &    &   &    &     &        &    & \OK        & & &     &        &   &  & 1     \\
   \textbf{Sonyc-UST} \cite{cartwrightSONYCUrbanSound2019}      & &      & &   & &&   &     & &     &       &   &      &  &    &   &    &     &        &    & \OK        & & &     &        &   &  & 1     \\
   \textbf{Zapsplat} \cite{DownloadFREESound}   & \OK     &      & &   & &&   &     & &     &       &   &      &  &    &   &    &     &        &    &   & & &     &        &   &  & 1     \\
   \textbf{Sound Jay} \cite{FreeSoundEffects} & \OK     &      & &   & &&   &     & &     &       &   &      &  &    &   &    &     &        &    &   & & &     &        &   &  & 1     \\
   \textbf{Pro Sound Effects} \cite{priceProSoundEffects}  & & \OK  & &   & &&   &     & &     &       &   &      &  &    &   &    &     &        &    &   & & &     &        &   &  & 1     \\
   \textbf{YouTube}  & & \OK  & &   & &&   &     & &     &       &   &      &  &    &   &    &     &        &    &   & & &     &        &   &  & 1     \\ \hline
   Total captions (M)*        & 0.1     & 50   & 0.1     &   & 0.6   & 3.3   &     & 1.2       & 2.4 & 5.6& 3.4      & 3.5    & 0.9        & 1.9 & 2.6     &  & 0.6   &    & 0.4& 0.4     & 5.9    & 3.9   & 3   & 0.9& 5.5& 0.8 & 0.1    &  \\ \hline
   Total hours (K)*  & & 145.6& & 4 & 4.3   & & 3.4      &     & 8       &     &       &   & 3.7  & 5.7&    & 2 &    & 6.5 & 1      & 2  &   & 1.3    & & 3.7 &        & 8.8      & 0.1     &\\ \bottomrule
   \end{tabular}
   }
   \end{table*}

\subsection{Datasets for Large Audio-Language models}
Current state-of-the-art audio-language models are typically trained on combinations of multiple datasets, which are described in this section. Table \ref{tab:alm-overview} provides a comprehensive overview of datasets used across major audio-language models. The early SoundWords model \cite{kilgourTextdrivenSeparationArbitrary2022} from Google utilized an exceptionally large training set of 50 million captions, while most subsequent models work with 1-5 million captions. The most widely adopted datasets are AudioCaps and its parent dataset AudioSet. Out of the 27 audio-language models analyzed, 21 used AudioCaps. The next most popular dataset was Clotho \cite{drossosClothoAudioCaptioning2020}, followed by various Freesound-derived datasets. Sound effect libraries like BBC SFX, Sonniss, and Audiostock are also frequently used in audio-language models for training.
\section{Audio Dataset Evaluation} 
In this section, an evaluation of the primary audio datasets is presented. First, visualizations are provided based on a large-scale principal component analysis of audio- and text-embeddings. These visualizations illustrate the relative distance between audio- and text- content in the various datasets and their absolute numbers in terms of AudioSet top-level categories.
Second, a quantitative evaluation of data leakage is presented. For this, the cosine similarity between CLAP embeddings of the audio and text data is used.

Given the large scale of the dataset, the calculation of the CLAP embeddings, the calculation of the PCA in section \ref{sec:audio-dataset-analysis} and the calculation of pairwise cosine similarity in section \ref{sec:dupe-analysis} presented significant computational challenges. The calculation of the CLAP embeddings using the laion/larger\_clap\_general model took approximately 2 days on a single H100 GPU. Across all datasets, 18,822,210 audio-text pairs were processed. After excluding datasets that directly overlap with or are derived from other datasets (e.g., Clotho-Detail being based on Clotho), 6,903,631 audio-text pairs were retained. After experimentation, the Rapids cuML GPU implementation of PCA \cite{teamRAPIDSLibrariesEnd2023} was utilized to efficiently reduce dimensionality. Similarly, pairwise cosine similarity computations for the whole dataset were performed in batches using PyTorch on GPU.\cite{paszkeAutomaticDifferentiationPyTorch2017}.

\subsection{Audio Dataset Analysis} \label{sec:audio-dataset-analysis}
\begin{figure*}[ht]
        \includegraphics[width=\textwidth]{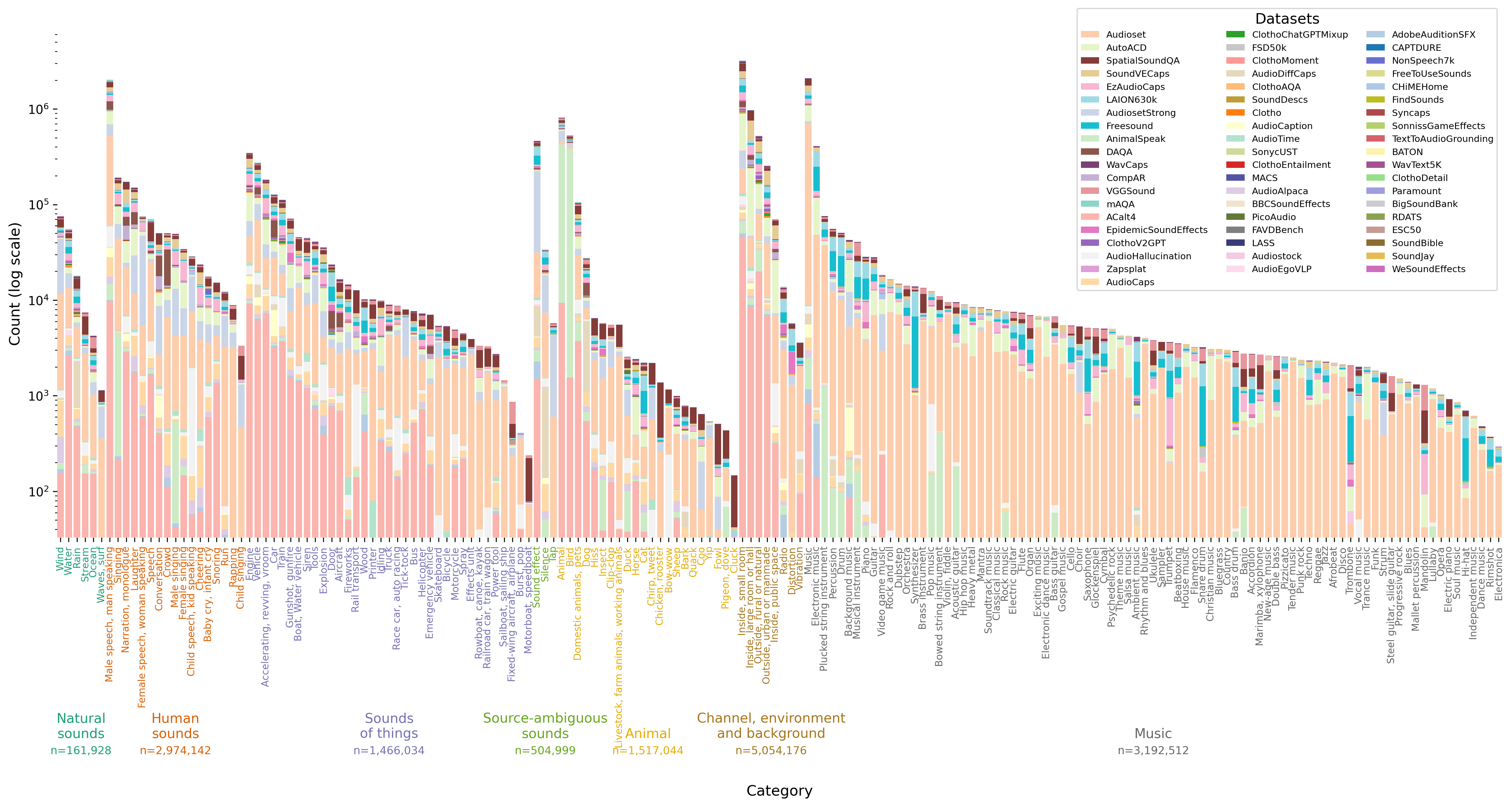}
        \begin{center}
        \caption{Distribution of audio samples across AudioSet's top-200 categories for each dataset stacked vertically, shown on a logarithmic scale. Categories are grouped by the root-level categories of the AudioSet ontology (Natural sounds, Human sounds, etc.), sorted by size and color-coded, with the total number of samples (n) shown below each group. Datasets are sorted by size in the legend.}
        \label{fig:clap-categories}
        \end{center}
 \end{figure*}
\subsubsection{Category Distribution}
In the first set of visualizations (Fig. \ref{fig:clap-categories}), audios and texts are categorized according to AudioSet's top-200 categories. This categorization is used to highlight the relative distribution of sounds in terms of general semantics. For this purpose, each text corresponding to a sound was processed using Meta's Llama 3.2 model (specifically, the \textit{meta-llama/Llama-3.2-3B-Instruct} model hosted on HuggingFace). The model was prompted with each caption and instructed to classify it into exactly one AudioSet category, returning only the exact full category name. The prompts were processed in batches of 64 and took approximately a week on a single H100 GPU. Samples were omitted from the visualization that could not be matched to a valid AudioSet category, which occurred in 3.8\% of the cases.

The visualization (Fig. \ref{fig:clap-categories}) shows the top 200 categories grouped by the 7 root-level categories of the AudioSet ontology: "Natural sounds", "Human sounds", "Sounds of things", "Source-ambiguous sounds", "Animal", "Channel, environment and background", and "Music". Among these root categories, only "Animal" and "Music" have their own direct class, while the others serve as grouping categories. The visualization is limited to the top 200 categories rather than all 527 AudioSet categories to avoid visual clutter. As a consequence, categories with less than 2,264 samples were excluded. Although the LLM was given 200 categories to choose from in its classification task, it only returned a total of 177 unique categories in all classifications. 

Comparing the model's classifications to AudioSet's original labels reveals some discrepancies. For example, "Male speech, man speaking" was classified 30.6 times more frequently by the model than it appears on AudioSet's labels, "Sound effect" 14.5 times more frequently, and "Inside, small room" 10.6 times more frequently. Some examples in the top 200 categories that are not included in the visualization are "Drum kit" with 15169 audios in AudioSet, "Keyboard (musical)" (10473) and "Motor vehicle (road)" (9044).

Based on the analysis of category distributions across audio datasets in Figure \ref{fig:clap-categories}, significant imbalances in category representation is observed. The most prominent categories show a strong bias towards indoor environments and human-generated sounds, with "Inside, small room" being heavily overrepresented at 21.6\% (3.2M samples), followed by "Music" and "Male speech" at 14.1\% and 13.6\% respectively. These three categories alone account for nearly 50\% of all samples, while specific sound categories like "Cluck" (147 samples), "Rimshot" (371 samples), and "Motorboat" (239 samples) are severely underrepresented. 

The analysis also shows notable patterns in category concentration within individual datasets. For instance, DAQA exhibits a high concentration of male speech (37.8\% of its samples), while BBCSoundEffects shows a strong bias toward indoor environments (34.2\% "Inside, small room" samples). In general, indoor environmental sounds and human-generated content dominate, while natural sounds and specific sound effects form a long tail in the distribution. Despite the "Animal" and "Bird" categories appearing in the top 10 most common categories (5.5\% and 3.5\% respectively), most nature-related and specific sound effect categories have minimal representation. The category distribution uniformity is also calculated using the formula:

\begin{align}
H_n(\alpha) = \frac{-\sum_{i=1}^n p_i \log p_i}{\log n} \\
\text{where } p_i = \frac{\alpha_i}{\sum_{j=1}^n \alpha_j}
\end{align}

Where $n$ is the number of categories, $\alpha_i$ represents the count in category $i$, and $p_i$ is the probability of category $i$. The normalized entropy ranges from 0 (completely concentrated in one category) to 1 (uniformly distributed across categories). A higher value indicates a more uniform distribution of categories within a dataset, something a balanced dataset should have. In this calculation, it is seen that VGGSound shows the most uniform distribution (0.720), followed by FindSounds (0.615) and BATON (0.610). This suggests that these datasets have more balanced representation across their categories; however, the fact that these datasets have a limited number of unique words in their captions influences this observation as well. When examining samples per category, SpatialSoundQA leads with 12,382 samples per category on average, followed by AudioSet (11,773) and DAQA (10,968). 

\subsubsection{PCA Embeddings}
\begin{figure*}[ht]
        \includegraphics[width=\textwidth]{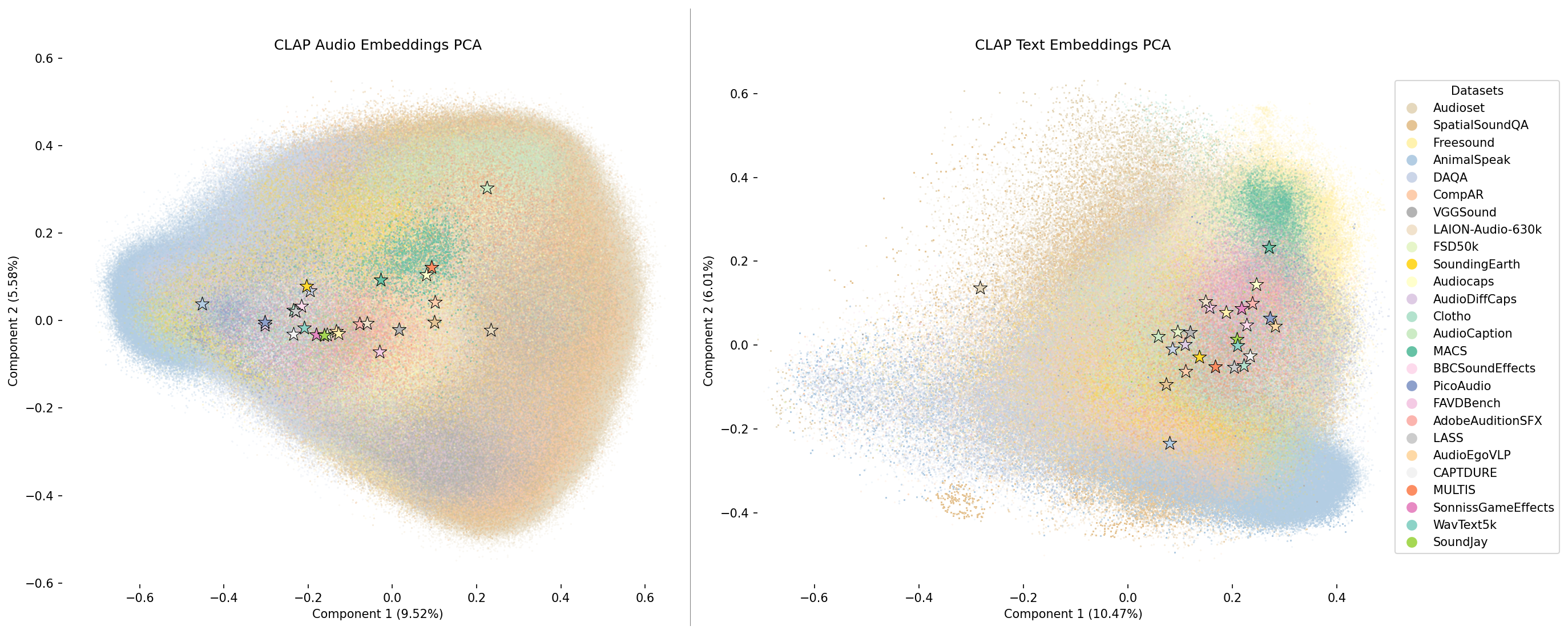}
        \begin{center}
        \caption{PCA decomposition of all the datasets on both the audio and the language. The stars in the visualisations represent the centroid of the dataset. } 
        \label{fig:pca-clap-embeddings}

        \end{center}
\end{figure*}

For the visualization of embeddings, only the original datasets are utilized (i.e. the ones without a parent dataset; see also the indentation of the datasets in Table \ref{tab:datasets}), with a few exceptions. For example, ClothoAQA is based on Clotho, and is therefore omitted from this analysis, since this dataset is already overlapping. Similarly, Auto-ACD, ESC50 and PicoAudio are also omitted. In addition, including all derivatives would make the visualizations too cluttered and difficult to interpret meaningfully. Datasets without available data, indicated with empty fields in Table \ref{tab:datasets}, are also omitted from this analysis. In total, 26 datasets are analyzed; visualizations of CLAP embeddings for both audio and text are generated (Fig. \ref{fig:pca-clap-embeddings}).

This set of visualizations (Fig. \ref{fig:pca-clap-embeddings}) represents the first and second principal components of audio (left) and text (right) embeddings obtained using the \textit{laion/larger\_clap\_general} CLAP model hosted on HuggingFace. In this set, the data points are color-coded based on the datasets (see legend) to highlight the acoustic and linguistic variability of each dataset and their centroids. These centroids are indicated by star glyphs and calculated by averaging all CLAP embeddings for that dataset after reducing them to two dimensions. For quantitative variability analysis, distance scores are calculated based on the pairwise Euclidean distance between each individual point in the datasets in the original embedding space. For further analysis, see also Section \ref{sec:dupe-analysis}.

In the audio embeddings PCA (Fig. \ref{fig:pca-clap-embeddings}, left), it can be observed that the centroid of the datasets AnimalSpeak, AudioCaption, MACS and AudioSet are the most distant from centroids of the other datasets. In the text embeddings PCA (right), the centroid of AudioSet is the most distant from the other centroids, which is likely due to the limited number of unique words in AudioSet's multi-label classification captions. Notably, the sounds of AnimalSpeak deviate in both audio and text embeddings, which may be because the dataset contains a large number of different animal sounds, which are, to a lesser extent, present in the other datasets.

In the original dimension, before conversion to 2 dimensions with PCA, Freesound (variability: 0.925), FSD50k (0.909) and WavText5k (0.906) show the highest variability in audio embeddings, while AnimalSpeak (0.699), AudioCaption (0.661) and MACS (0.534) show the lowest variability in audio embeddings. This variability is defined by the average distance for all embeddings to its centroid. For the text embeddings, Freesound (0.930), LAION-Audio-630k (0.927) and SonnissGameEffects (0.910) show the highest variability, while AudioCaption (0.688), AudioDiffCaps (0.674) and MACS (0.598) show the lowest variability. Freesound is a large collection of user-uploaded sounds, containing a large number of different sounds, which may explain its high variability in both audio and text embeddings. MACS on the other hand, represents sounds from four different locations (airport, public square, and park) and is therefore more homogeneous.

\subsection{Duplicate analysis} \label{sec:dupe-analysis}

\begin{figure*}[ht]
        \includegraphics[width=\textwidth]{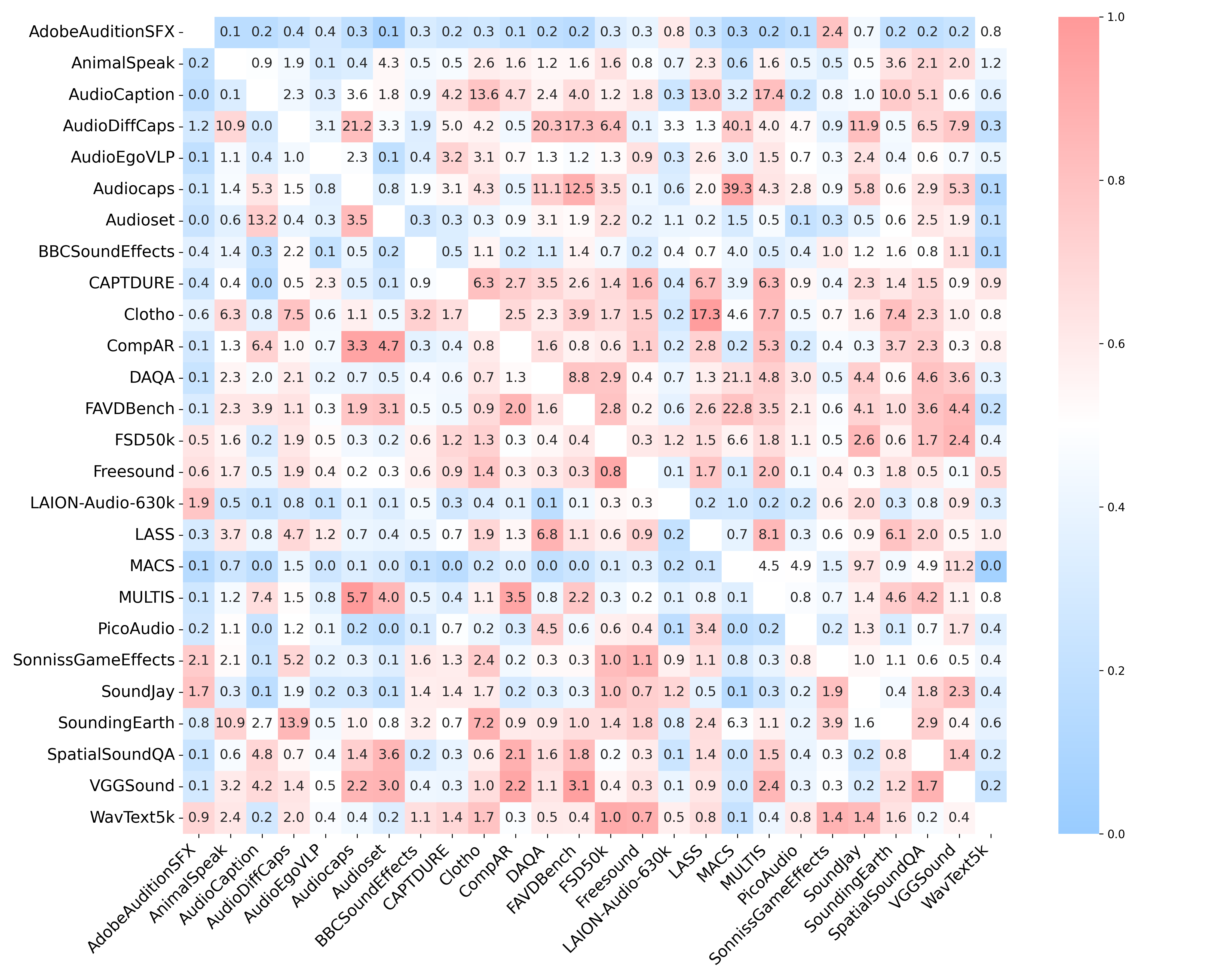}
        \begin{center}
        \caption{Heatmap showing the centroid distances and pairwise similarities between audio and text CLAP embeddings of all datasets. The color intensity represents the centroid distance, while the numerical annotations show the percentage of pairs exceeding 50\% similarity threshold between datasets. The lower (left) triangle represents audio similarities, and the upper (right) triangle represents text similarities.}
        \label{fig:heatmap-dupes}
        \end{center}
 \end{figure*}

Previous research has highlighted the issue of duplicate data in multimodal models, where pretraining datasets often include examples that the model is later expected to predict in a zero-shot context. This overlap can lead to inadequate evaluation of the models' generalization capabilities \cite{udandaraoNoZeroShotExponential2024}. Additionally, studies have shown that a significant portion of the zero-shot learning accuracy in audio-language models is attributable to the  inherent capabilities of their audio and text backbone models, rather than a genuine ability to perform zero-shot learning.  \cite{tavaresClassSeparabilityPitfalls2024}.

Several studies in audio-language datasets have highlighted dataset overlaps, which are particularly relevant for zero-shot inference. For example, when training models on WavCaps \cite{meiWavCapsChatGPTAssistedWeaklyLabelled2023}, researchers excluded data from FreeSound and AudioSet to ensure zero-shot evaluation on Clotho and AudioCaps, due to overlapping samples. Similarly, in training LAION-CLAP \cite{wuLargeScaleContrastiveLanguageAudio2023}, samples from various sound event datasets were removed to avoid overlap with Clotho and between the AudioCaps test set and the AudioSet training set. In the HEAR challenge \cite{turianHEARHolisticEvaluation2022}, datasets such as FSD50k and Mridingham were excluded from the benchmark to prevent overlap with Freesound,  \cite{zhuCacophonyImprovedContrastive2024}. The DCASE challenge 2024 published a CSV file of forbidden Freesound sound IDs to avoid the overlap between Freesound and Clotho. 

The duplicate analysis is inspired by two key papers on specific audio-language datasets. \citet{braliosGenerationReplicationAuscultating2023} examined memorization in latent diffusion models, identifying numerous duplicated audio clips in the AudioCaps database using CLAP embeddings and mel-spectrogram analysis. Similarly, \citet{weckDataLeakageCrossmodal2023} analyzed the SoundDescs dataset \cite{koepkeAudioRetrievalNatural2022}, detecting duplicates or overlapping recordings and producing a revised version free of redundancy by applying an audio fingerprinting algorithm \cite{sixPanakoScalableAcoustic2014}. However, an attempt to use this CPU-based, single-threaded $O(n^2)$ approach on about 20\% of the datasets in this survay took approximately a month, rendering it impractical for broader application.

To examine dataset similarities, we generated a heatmap visualization that combines both centroid and pairwise similarities for audio and text embeddings (Fig. \ref{fig:heatmap-dupes}). The computation is performed using GPU-accelerated libraries (cuPy and PyTorch) for efficiency. Centroid similarities were calculating using  mean embedding vectors for each dataset, with pairwise cosine similarities determining their relationships. In addition, pairwise similarities were computed for all embedding pairs and thresholds, of 50\% and 99\% similarity were used to quantify relationships.  The similarity threshold of 50\% is used in the visualization to highlight highly similar datasets. The diagonal is masked, since self-similarities are not meaningful for the analysis. 
The 99\% similarity threshold is used to define non-overlapping subsets of datasets, which are provided on our GitHub repository.

In the heatmap, colors represent centroid similarity scores ranging from 0 to 1, where higher scores indicate closer relationships. For pairwise similarities, the percentage of pairs exceeding the 50\% threshold provides additional insight into dataset overlap. In some cases, higher relative percentages may correspond to a subset of one dataset being highly similar to another, while overall similarity remains low. Conversely, a lower percentage with a high centroid score might indicate broader similarities across datasets.

Analysis of the audio centroid distances in the heatmap's bottom-left triangle (Fig. \ref{fig:heatmap-dupes}), reveals strong acoustic overlap in pairs like AudioCaps-MULTIS (similarity: 0.993), FAVDBench-VGGSound (0.964), CompAR-AudioSet (0.942), AudioCaps-CompAR (0.942), and MULTIS-CompAR (0.939). For some of these pairs, the overlap can be explained by known relationships - for instance, MULTIS and CompAR are both build upon AudioSet, which also includes AudioCaps samples. In contrast, FAVDBench's overlap is less clear but may originate from the inclusion of YouTube sources, which may contain VGGSound samples \cite{shenFinegrainedAudibleVideo2023}. Scores close to 0 indicate that datasets exhibit minimal acoustic overlap, suggesting that their combination may be beneficial in training audio-language models. In particular, the lowest scores are found between DAQA-LAION-Audio-630k (0.176), MACS-CAPTDURE (0.167), CAPTDURE-AudioCaption (0.165), SoundJay-MACS (0.148), and AdobeAuditionSFX-MACS (0.148). 

The heatmap's upper-right triangle (Fig. \ref{fig:heatmap-dupes}) shows high text similarity, as assessed through the centroids, in pairs like LASS-Clotho (similarity: 0.980), MACS-AudioCaps (0.935), FAVDBench-AudioCaps (0.887), VGGSound-FSD50k (0.859), and LASS-MULTIS (0.847), reflecting shared vocabulary and captions. Meanwhile, the lowest similarities occur between AdobeAuditionSFX-MULTIS (similarity: 0.148), AdobeAuditionSFX-MACS (0.146), BBCSoundEffects-WavText5k (0.144), AudioCaps-WavText5k (0.137), and WavText5k-MACS (0.056). Overall, AdobeAuditionSFX (average similarity: 0.280) and WavText5k (0.375) have the lowest text similarity scores to all other datasets. FSD50k (0.631) and SpatialSoundQA (0.623). For audio, this is MACS (0.273) and AdobeAuditionSFX (0.376), and Freesound (0.629), VGGSound (0.606).

Analysis of pairwise audio embedding similarities, as indicated by the numerical value in each cell, shows that AudioDiffCaps-SoundingEarth (266M pairs, 13.89\%), AudioSet-AudioCaption (5.9B pairs, 13.24\%), and AnimalSpeak-SoundingEarth (4.8B pairs, 10.93\%) have the highest number of pairs exceeding the 50\% similarity threshold. For the 99\% similarity threshold, SonnissGameEffects-VGGSound (22,880 pairs, 0.21\%), MULTIS-AudioCaps (5,044 pairs, 0.16\%), and Freesound-SonnissGameEffects (29,016 pairs, 0.05\%) show the highest number of relative similar pairs. The lowest number of relative pairs exceeding the 50\% similarity threshold are found between MACS-PicoAudio (11,918 pairs out of 258M total comparisons, 0.0046\%), PicoAudio-AudioCaption (17,280 pairs out of 286M, 0.0060\%), and CAPTDURE-AudioCaption (11,390 pairs out of 186M, 0.0061\%).

Overall, this analysis reveals interesting patterns across datasets. AudioDiffCaps shows high acoustic similarity with several datasets, particularly SoundingEarth (13.89\% of pairs >50\% similar), AnimalSpeak (10.88\%), and Clotho (7.53\%). Similarly, AudioCaption exhibits strong acoustic relationships with multiple datasets, having over 5\% similarity with MULTIS (7.35\%), CompAR (6.40\%), and AudioSet (13.24\%). In contrast, specialized datasets like MACS and AdobeAuditionSFX show minimal acoustic overlap - MACS has less than 0.5\% similarity with all other datasets except SoundingEarth (6.29\%), while AdobeAuditionSFX's highest similarity is only 1.94\% with LAION-Audio-630k. This suggests these datasets contain unique acoustic content not well-represented in other datasets.

The analysis of pairwise text embedding similarities shows that AudioDiffCaps-MACS (312.8M pairs, 40.07\%), Audiocaps-MACS (335.1M pairs, 39.31\%), and FAVDBench-MACS (45.8M pairs, 22.80\%) have the highest number of pairs exceeding the 50\% similarity threshold. For the 99\% similarity threshold, SoundJay-MACS (420 pairs, 0.0018\%), SoundJay-VGGSound (1,502 pairs, 0.00068\%), and Audiocaps-VGGSound (37,199 pairs, 0.00046\%) show the highest number of relative similar pairs. The lowest number of relative pairs exceeding the 50\% similarity threshold are found between MACS-WavText5k (44,783 pairs out of 94.2M total comparisons, 0.048\%), AnimalSpeak-AudioEgoVLP (4.2M pairs out of 7.7B, 0.054\%), and AudioEgoVLP-AudioSet (9.9M pairs out of 17.5B, 0.057\%).

In general, datasets like WavText5k consistently show low text similarity with other datasets, with its highest similarity being only 1.16\% with AudioCaption. Similarly, AdobeAuditionSFX shows limited textual overlap, with its maximum similarity being 2.39\% with SonnissGameEffects. On the other hand, AudioDiffCaps again shows high textual similarity with several datasets, particularly AudioSet (21.2\%), DAQA (20.3\%), and FAVDBench (17.3\%). Particularly noticeable are the 39.3\% and the 40.1\% similarities in text between LASS and AudioCaps and AudioDiffCaps respectively.

For all datasets, our GitHub repository provides subsets of datasets that do not overlap with other data, based on the 99\% threshold. The GitHub repository also contains instructions and easy-to-use bash scripts for downloading each dataset in our analysis.

\section{Discussion} \label{sec:discussion}

\subsection{Data quality}
The quality of training data significantly influences the performance of machine learning models. While high-quality training datasets are limited in size \cite{zhuCacophonyImprovedContrastive2024}, relying on large-scale, noisy datasets remains the only scalable option for many tasks. \cite{elizaldeClapLearningAudio2023}. However, as noted in the AudioCaps paper, such datasets may induce distributional shifts \cite{kimAudioCapsGeneratingCaptions2019}, particularly in generating audio samples with user-generated free-form text inputs instead of training set prompts \cite{changContextPromptEditing2023}. \citet{meiWavCapsChatGPTAssistedWeaklyLabelled2023} showed that filtering noisy data and rewriting raw descriptions into high-quality captions can achieve state-of-the-art performance with smaller datasets. Similarly, augmenting AudioCaps with ChatGPT resulted in an enhanced performance on audio-text retrieval \cite{ohDistanceSamplingbasedParaphraser2024}.

The combination of datasets poses additional challanges. For instance, \citet{meiWavCapsChatGPTAssistedWeaklyLabelled2023} observed a performance drop in the AudioCaps dataset when the LAION-Audio-630K dataset, which is larger but presumably of lower quality, was included during training. However,  integrating the WavCaps dataset reversed this decline and led to significant improvements in the Clotho and AudioCaps datasets despite the smaller dataset size. These findings emphasize the importance of data quality over quantity.

LLMs offer a promising solution to enhance data set quality. Rewriting existing datasets to produce more elaborate and diverse captions, has been shown to improve training outcomes and mitigate some of the adverse effects of noisy data.\cite{wuImprovingAudioCaptioning2024, primusAdvancingNaturalLanguageBased2023, xiaoEnsembleSystemsContrastive2023, yuanSoundVECapsImprovingAudio2024, kongImprovingTextAudioModels2024}. A key challenge, however, is the inconsistency in caption style across audio-language datasets \cite{zhuCacophonyImprovedContrastive2024}. Proposed solutions include introducing dataset-specific tokens \cite{labbeCoNeTTEEfficientAudio2023} or standardizing caption style and length using LLM-assisted rewriting (e.g. \cite{kimOvercomingDataShortage2023,ghoshRecapRetrievalaugmentedAudio2024, sunAutoACDLargescaleDataset2024}).

\subsection{Bias in Audio Datasets}
Audio-language models face various biases that can affect their fairness and performance. Key sources of bias include crosscontamination of train/test splits, the bias related to the visual-grounding of datasets, diversity bias and the bias related to dataset augmentation (model collapse).

\subsubsection{Cross-contamination} 
Cross-contamination occurs when training and testing datasets overlap, leading to overly optimistic performance metrics, as the model is essentially being tested on information it has already seen and masking a model's generalizability. This issue is prevalent in the context of Audio-Language Learning, especially when combining  datasets.  For example, 17.6\% of AudioCaps overlaps with WavCaps, while Clotho shares an even larger overlap of 89\% \cite{labbeCoNeTTEEfficientAudio2023}. Such overlaps distort evaluations and undermine real-world applicability. The dataset overlap analyses provided in this survey can help avoiding such cross-contamination. 

\subsubsection{Visually-grounded datasets} 
Many prominent audio-language datasets, such as AudioCaps, were annotated with access to corresponding video content\cite{kimAudioCapsGeneratingCaptions2019} (see Section \ref{sec:datasets} and Appendix \ref{app:audio-visual-datasets}). This introduces a risk of visual bias, where captions include information not discernible from audio alone (e.g., "A female instructor giving a speech in front of a live audience"). \citet{xuBLATBootstrappingLanguageAudio2023} found that noise introduced by visual context limits improvements in audio-text alignment studies \cite{yanpengzhaoConnectingDotsAudio2022, guzhovAudioCLIPExtendingCLIP2021, wuWav2CLIPLearningRobust2022}. Efforts to address this include filtering visually-influenced captions using LLM (GPT4o), though such method resulted in an audio-language dataset that contained only 0.65\% of the original data (70,000 of 10.8 million) \cite{kulikTakeItGranted2024}. 

Several datasets such as Auto-ACD \cite{sunAutoACDLargescaleDataset2024} and Sound-VECaps \cite{yuanSoundVECapsImprovingAudio2024} have been created by augmenting audio datasets with visual information to enrich their content. Research by \cite{nishimuraAudioHallucinationsLarge2024} showed however that visual language models tasked with processing both video and audio tend to generate audio descriptions predominantly based on visual information rather than the actual audio content. Their analysis of 1000 cases showed that Video-LLaMA produced audio descriptions containing visual-only information in 32.3\% of cases. To mitigate this, several works proposed to filter out visual-only information from the caption \cite{yuanSoundVECapsImprovingAudio2024, sunAutoACDLargescaleDataset2024, haiEzAudioEnhancingTextaudio2024}.

\subsubsection{Diversity bias}
Audio-language models trained on unbalanced, non-diverse datasets may perform poorly on underrepresented groups and cultures \cite{romanDiscreteTokensHighfidelity2023a}. Similarly, audio retrieval systems may exhibit lower accuracy when searching for content related to cultures or languages poorly represented in the dataset. Most existing datasets focus on English, with only a few exceptions, such as AudioCaption \cite{wuAudioCaptionListen2019, xuAudioCaptionCar2021} and FAVDBench \cite{shenFinegrainedAudibleVideo2023}, both having original Mandarin captions translated to English, and CAPTDURE, translated from Japanese to English \cite{okamotoCAPTDURECaptionedSound2023}. Efforts to expand linguistic diversity include translating datasets into multiple languages. \citet{cousinMultilingualAudioCaptioning2023} used automatic translation to translate AudioCaps and Clotho to German, French, and Spanish. \cite{manakulEnhancingLowResourceLanguage2024a} translated AudioCaps from English to Thai. For audio question answering, \cite{beheraMultiLingualAudioQuestion2023} translated into ClothoAQA in 8 languages. \citet{kangDevelopmentKoreanAudio2020} build a Korean audio captioning system, by first training the system in English and then translating the captions from English to Korean. This approach has also been scaled to 181 languages by \citet{noriyCLARAMultilingualContrastive2023}. The authors found that this multilingual data augmentation enriches the variability of languages and accents within their training set, enhancing the model's generalization capabilities. However, relying on automatic translation can introduce its own bias. Annotator proficiency in dataset languages also raises concerns, particularly when using platforms like Mechanical Turk, known for demographic biases \cite{hekanahoLanguagebasedMachinePerception2024, difallahDemographicsDynamicsMechanical2018}.

\subsubsection{Model collapse}
Augmentation techniques have become increasingly important in audio-language research to address data scarcity and improve model performance. Research has shown that the use of ChatGPT and other LLMs for text augmentation leads to higher-quality datasets for training of audio language models through filtering captions \cite{kreukAudioGenTextuallyGuided2023, robinsonTransferableModelsBioacoustics2024, xiaoEnsembleSystemsContrastive2023} or generating new captions \cite{robinsonTransferableModelsBioacoustics2024, primusAdvancingNaturalLanguageBased2023,kimOvercomingDataShortage2023, sunAutoACDLargescaleDataset2024, ghoshRecapRetrievalaugmentedAudio2024, guLanguagebasedAudioRetrieval2024}. On the other end, attempts to augment audio data using text-to-audio generation models have shown limited success due to insufficient diversity and realism in generated samples \cite{liuWavJourneyCompositionalAudio2023, ghosalTextaudioGenerationUsing2023, ghoshCompAAddressingGap2024}. For example, \cite{ghoshCompAAddressingGap2024} attempted to create compositional audio using cutting-edge text-to-audio generation models \cite{liuWavJourneyCompositionalAudio2023, liuAudioLDMTexttoAudioGeneration2023, ghosalTextaudioGenerationUsing2023} to enhance training for audio-language models, but found limitations in generating diverse and cohesive audio for acoustic scenarios and concluded that a significant gap in the capability of current text-to-audio models is still present. 
This limited success is a consequence of "model collapse" \cite{shumailovAIModelsCollapse2024}, a phenomenon common in many areas of machine learning where models relying on synthetic data perform well on such data but fail to generalize to real-world scenarios.

\subsection{Dataset Accessibility}
Dataset accessibility and reproducibility remain significant challenges in audio-language research. Many datasets, such as those derived from AudioCaps, require researchers to scrape audio content manually using command-line tools like \textit{yt-dlp}, or \textit{yt-dl} for YouTube-based content. Some others follow potentially cumbersome naming conventions \cite{kongPANNsLargeScalePretrained2020},such as adding a 'Y' prefix to filenames - a practice seen in \cite{kongPANNsLargeScalePretrained2020} and subsequent works - to handle YouTube IDs that might contain symbols problematic for programmatic parsing. For other datasets, such as AnimalSpeak, SoundingEarth, and Zapsplat, obtaining the data is a time-consuming and complex process. Instructions for accessing datasets like DAQA and AudioEgoVLP are often unclear, while others, such as SoundScaper, present barriers so significant that they hinder practical use.

Reusing existing datasets as building blocks for new ones further complicates access and reproducibility, as acquiring a single dataset may necessitate downloading multiple prerequisites. To improve this situation, researchers should prioritize hosting datasets on accessible platforms such as Zenodo or HuggingFace, ensuring straightforward access and facilitating reproducibility. Our GitHub repository provides scripts for easy acquisition of datasets.
\section{Conclusion} \label{sec:conclusion}
This survey provides a comprehensive overview of audio-language datasets and their role in developing audio-language models. The field has seen significant progress due to advances in computing power and the creation of large-scale datasets like AudioCaps and Clotho. This enabled the training of more sophisticated models capable of understanding and describing diverse sounds. This survey examined various aspects, including dataset design, availability, training techniques, and challenges in the field.

A key finding is the increasing use of language models for dataset augmentation and caption generation. Currently, 35 out of the 69 analyzed datasets have used LLMs to generate captions for their datasets. While LLMs can help tailor datasets for specific model purposes, our analysis reveals potential redundancy as many LLM-generated variations of the same datasets may already exist (e.g. several similar versions of Clotho). The section \ref{sec:discussion} highlighted how this promising approach needs careful consideration. Both bias and hallucination issues could occur when using LLMs to generate captions.

This survey emphasizes the role of data quality and quantity, particularly for contrastive learning approaches such as CLAP. While larger datasets enable better model performance, the literature shows that simply increasing dataset size is not sufficient. In this regard, the quality and diversity of the data are equally important. Studies show that incorporating lower-quality datasets can actually decrease model performance, while smaller but higher-quality datasets can lead to improvements.

The analysis in Table 1 reveals that YouTube-based datasets dominate the field (23 datasets), followed by Freesound-based datasets (21 datasets). Sound effect libraries and field recording datasets also form a large portion of datasets (18 and 15 datasets, respectively). However,sound effect libraries sometimes require payment or special access for their use. Other datasets, particularly AudioSet-derivatives, are difficult to access since AudioSet's YouTube-based constraints prevent direct resharing of audio data, requiring a manual download. Interestingly, we found that 49.2\% (34 out of 69) of datasets use LLM-generated captions, indicating a clear trend toward synthetic data generation. This trend is particularly strong in recent years - in 2024, 80\% of new datasets (20 out of 25 of datasets released that year) used LLM-generated captions, up from 69.2\% (9 out of 13) in 2023.

The analysis of sound content distributions in the audio datasets reveals significant imbalances in sound category representation, with indoor environments and human-generated sounds dominating the landscape. Our analysis shows that "Inside, small room" (21.6\%), "Music" (14.1\%), and "Male speech" (13.6\%) account for nearly 50\% of all samples, while specific sound categories remain severely underrepresented. Dataset-specific patterns emerged, such as DAQA's high concentration of male speech (37.8\%) and BBCSoundEffects' bias toward indoor environments (34.2\%). When examining dataset uniformity, VGGSound showed the most balanced distribution (0.720 normalized entropy), followed by FindSounds (0.615) and BATON (0.610). PCA analysis of CLAP embeddings revealed that AnimalSpeak, AudioCaption, MACS, and AudioSet had the most distinct centroids. At the same time, Freesound demonstrated the highest variability in both audio (0.925) and text (0.930) embeddings, reflecting its diverse user-uploaded content.

Duplicate analysis as shown in Section \ref{sec:dupe-analysis} revealed significant overlap between various datasets. It is important to be aware of data leakage when training and evaluating your models, particularly when performing zero-shot learning evaluation. The heatmap analysis showed great similarities between certain pairs of datasets, such as VGGSound-FAVDBench and Freesound-FSD50k. Large audio-language models may have seen similar data during pretraining, which can lead to over-optimistic performance metrics. It is not recommended to perform strong zero-shot evaluation on datasets that share similar origins (e.g., training on Clotho and evaluating on FSD50k, or training on Auto-ACD and evaluating on AudioCaps). This evaluation is better described as a weak zero-shot evaluation. 

Some overlaps are more obvious than others. The following are some examples of overlaps: VGGSound and AudioSet overlap. ProSoundEffects and BBC Sound Effects also have overlap, as Pro Sound Effects operate the commercial licensing of BBC sound effects. Auto-ACD is based on VGGSound and AudioSet, WavCaps is based on FreeSound, BBC, SoundBible and AudioSet. AnimalSpeak further contains also AudioCaps as a subset. The analysis in this survey also deduced that the overlap between VGGSound and FAVdBench Audio is quite evident. For each dataset, specific datasets without overlaps are provided in our GitHub repository.

A relevant trend is the increasing amount of different tasks and subtasks that the field of Audio-Language Learning is exploring. This survey shows that 8 different tasks are currently being explored, with a further division of 11 subtasks. As Audio-Language models become more usable for real-world applications, it is likely that more tasks will be explored. In particular, various tasks that  originally involved sound classification are now being extended to broader language involvement, such as text-queried sound event localization, language-queried audio source separation, and audio difference captioning.

The analysis of dataset statistics reveals varying approaches to caption length and detail. While some datasets focus on single sentence captions (e.g., AudioCaps with an average of 49.22 characters per caption or MACS with an average of 54.26), others contain more elaborate captions (e.g., ClothoDetail with an average of 324.83 characters per caption). This trend of more elaborate captions is likely to continue, particularly for instruction-tuning large audio-language models, as it allows for more detail and context to be captured in the training data.

\subsection{Future Research Directions and Recommendations}

Looking ahead, this survey identifies several areas for future dataset development. Most existing datasets are heavily skewed toward English-language content. While some datasets like AudioCaption and FAVDBench have original Mandarin captions and CAPTDURE has Japanese captions, the vast majority of audio-language datasets are English-only. Current approaches largely rely on automatic translation of English datasets to other languages, as seen in efforts to translate AudioCaps and Clotho to languages such as German, French, Spanish, and Thai. Although translation can help address immediate needs, it may not capture the full linguistic and cultural nuances of different languages. The development of original, high-quality datasets in diverse languages would be valuable to advance truly multilingual audio language models.

A significant gap also exists in specialized domains, with some noticeable exceptions: AnimalSpeak in the bioacoustics domain, AudioCaption Hospital in the medical domain, SoundScaper Dataset for soundscapes and MIMII-Change for malfunctions in industrial machines. This survey shows that most datasets focus on human activities, speech, and urban sounds, leaving room for expansion to other underexplored acoustic domains, such as environmental audio, niche industrial applications, and natural soundscapes.

YouTube-based datasets, while numerous, face challenges related to link rot and content availability, affecting research reproducibility. These findings suggest that future efforts should focus not only on increasing the size of the dataset but also on ensuring quality, accessibility, and robustness. Emphasis should be placed on reducing redundancy and addressing biases and representation gaps to ensure that datasets remain reliable and inclusive.

It is also important to address model collapse in audio-language datasets. While LLMs have shown promise in text augmentation and caption filtering, attempts to use audio generation models for data augmentation have had mixed results. Current text-to-audio generation models still show significant limitations in generating diverse and cohesive audio for acoustic scenarios. Future research needs to focus on developing more robust audio generation techniques that can produce high-quality synthetic data while avoiding the pitfalls of model collapse, where models over-rely on synthetic data and perform poorly on real-world examples.

To improve pretraining for audio-language models, researchers should explore alternative sources beyond Freesound-based datasets like Clotho or YouTube-based datasets such as AudioCaps. This can diversify the dataset and result in a more robust model. Some sound effect datasets and field recording datasets show the least overlap in this survey's analysis, offering promising alternatives for dataset diversification. WavText5k, BBC Sound Effects, Adobe Audition SFX and MACS are particularly recommended.

Dataset accessibility remains a pressing concern. Researchers should prioritize making their datasets directly available through established platforms like Zenodo or HuggingFace to ensure easier access and better reproducibility. Researchers should also be transparent about their dataset acquisition process to avoid confusion and ensure that their datasets are accessible to others.

Finally, this survey highlights the need for more sophisticated duplicate detection and dataset curation methods. Our analysis revealed complex patterns of overlap between datasets, with some similarities being more subtle than others. Although current methods, such as audio fingerprinting and embedding-based similarity detection, provide insights, they are often computationally intensive and limited in their ability to detect subtle overlaps. Future research should focus on developing more efficient and accurate methods for detecting duplicates and overlaps in audio-language datasets. This would enable better dataset curation and more reliable evaluation of zero-shot learning capabilities, while also addressing the current challenges in assessing true model generalization versus performance inherited from overlapping training data.

In conclusion, this survey serves as a valuable resource for researchers in audio-language modeling by providing detailed insights into the current state of datasets, their applications, and challenges. The analysis reveals trends in dataset development, including the growing use of LLMs for caption generation, the difference in ease of dataset acquisition and the issue of dataset overlap. These findings can guide researchers in making informed decisions about dataset selection and future development efforts in the field of audio-language learning.

\section*{Acknowledgment}
This work was supported by the Dutch Research Council (NWO 406.20.GO.030 to Prof. Elia Formisano), the Dutch national e-infrastructure with the support of the SURF Cooperative using grant no. EINF-6190, Data Science Research Infrastructure (DSRI; Maastricht University) and the Dutch Province of Limburg.

\bibliographystyle{IEEEtran}
\bibliography{library}

\appendices

\section{List of audio-visual datasets} \label{app:audio-visual-datasets}
\begin{itemize}
\item Video Captioning and Dense Description: ActivityNet-Captions \cite{krishnaDenseCaptioningEventsVideos2017}, MSVD \cite{chenCollectingHighlyParallel2011}, MSR-VTT \cite{xuMSRVTTLargeVideo2016}, VATEX \cite{wangVATEXLargeScaleHighQuality2020}, TaCos \cite{rohrbachCoherentMultiSentenceVideo2014}.
\item Instructional and Procedural Content Understanding: YouCook \cite{zhouAutomaticLearningProcedures2018}, HowTo100M \cite{miechHowTo100MLearningTextVideo2019}, CrossTask \cite{zhukovCrosstaskWeaklySupervised2019}, How2 \cite{sanabriaHow2LargescaleDataset2018}.
\item Visual Instruction Tuning: VideoChat-instruct-11K \cite{liVideoChatChatcentricVideo2023} and Valley-Instruct-65k \cite{luoValleyVideoAssistant2023}
\item Event Localization and Action Recognition: AVE \cite{tianAudioVisualEventLocalization2018}, TVC \cite{leiTVRLargeScaleDataset2020}, LLP \cite{tianUnifiedMultisensoryPerception2020}. DiDeMo \cite{hendricksLocalizingMomentsVideo2018}, Charades-STA \cite{gaoTALLTemporalActivity2017}. 
\item Large-Scale Multimodal Integration: HD-VILA-100M \cite{xueAdvancingHighResolutionVideoLanguage2022}, Webvid10M \cite{bainFrozenTimeJoint2021}, VIDAL-10M \cite{zhuLanguageBindExtendingVideoLanguage2024}, VALOR-1M \cite{chenValorVisionaudiolanguageOmniperception2023}, VAST-27M \cite{chenVASTVisionAudioSubtitleTextOmniModality2023}, ViT-Lens-2 \cite{leiViTLens2GatewayOmnimodal2023}.
\item Specialized Domains: LSMCDC/MPII-MD  \cite{rohrbachMovieDescription2017, rohrbachDatasetMovieDescription2015}, M-VAD \cite{torabiUsingDescriptiveVideo2015}, Sound Of Story \cite{baeSoundStoryMultimodal2023}, SoundNet \cite{aytarSoundNetLearningSound2016}, ASIW \cite{chatterjeeVisualSceneGraphs2021}
\end{itemize}
\section{Example caption per dataset}
\begin{table}[ht]
\caption{Example caption per dataset: the first occurrence of the word dog in each dataset.}
\label{app:dog-captions}
\scalebox{0.88}{
\begin{tabular}{@{}p{2.5cm}p{6.7cm}@{}}
\toprule
\textbf{Dataset} & \textbf{Example Caption}\\ \midrule
\textbf{ACalt4} & A dog whimpering and yipping, possibly in discomfort or seeking attention \\ \midrule
\textbf{AFAudioSet} & The audio contains a variety of sounds including a dog growling, the whirring sound of mechanisms, and a man speaking. \\ \midrule
\textbf{AdobeAuditionSFX} & Animal Mammal Carnivore Dog Bulldog Jowl Flap 01 \\ \midrule
\textbf{AnimalSpeak} & The sound of a Magnolia Warbler singing from dense dogwood foliage. \\ \midrule
\textbf{AudioAlpaca} & A female's speech followed by a dog barking and whimpering. \\ \midrule
\textbf{AudioCaps} & A dog whimpers quietly \\ \midrule
\textbf{AudioCaption} & A woman nurse and a female doctor talk about dogs. \\ \midrule
\textbf{AudioCondition} & Mechanisms from 0.0 to 10.0 and Generic impact sounds from 0.128 to 0.259 and Clicking from 0.467 to 0.543 and Whimper (dog) from  0.688 to 1.152 and Dog from 1.124 to 4.817 and Clicking from 4.775 to 4.824 and Surface contact from 4.886 to 5.066 and Dog from 5.149  to 6.912 and Clicking from 5.17 to 5.336 and Generic impact sounds from 5.979 to 6.186 and Generic impact sounds from 6.435 to 6.629 and  Generic impact sounds from 6.857 to 7.002 and Generic impact sounds from 7.327 to 7.5 and Human voice from 7.68 to 8.178 and Generic  impact sounds from 7.791 to 7.998 and Generic impact sounds from 8.288 to 8.489 and Generic impact sounds from 8.745 to 8.966 and Generic  impact sounds from 9.236 to 9.409 and Human voice from 9.277 to 9.838 and Generic impact sounds from 9.713 to 9.927 \\ \midrule
\textbf{AudioDiffCaps} & remove sound of dog barking \\ \midrule
\textbf{AudioEgoVLP} & The hotdog display counter is closed, producing a sliding and locking sound. \\ \midrule
\textbf{AudioHallucination} & Is there a sound of dog in the audio? No \\ \midrule
\textbf{AudioTime} & A dog barked for 1.385 seconds, followed by a steam whistle lasting 5.25 seconds. \\ \midrule
\textbf{AudioSet} & Speech, Whimper (dog) \\ \midrule
\textbf{AudioSetStrong} & Whimper (dog) \\ \midrule
\textbf{Audiostock} & Medium-sized dog barks (one!) \\ \midrule
\textbf{AutoACD} & A dog howls loudly followed by a person screaming, creating a chaotic atmosphere in the vicinity. \\ \midrule
\textbf{BATON} & A dog barks, bees buzz, and rail transport moves \\ \midrule
\textbf{BBCSoundEffects} & Ext. aerobatics as in dog fight - 1967 (104B \\ \midrule
\textbf{BigSoundBank} & barking dogs. \\ \midrule
\textbf{CAPTDURE} & The sound of a dog on a leash "roaming" on an inorganic floor. \\ \midrule
\textbf{CHiMEHome} & - \\ \midrule
\textbf{Clotho} & A dog jumped into the water and swim and then came out panting, it shook its body \\ \midrule
\textbf{ClothoAQA} & Is it a dog making the noise? no \\ \midrule
\textbf{ClothoChatGPTMixup} & a couple converses as a dog barks intermittently while coins are dropped into a jar and a hand searches through a packet \\ \midrule
\textbf{ClothoDetail} & In this audio, we hear a series of creaking and squeaking sounds. These noises could be attributed to someone leaning back and forth on an office chair, or perhaps to a slowly creaking door opening. Additionally, there are intermittent popping sounds that eventually taper off. Background noises, such as a dog barking, add additional context and atmosphere to the scene. \\ \midrule
\textbf{ClothoEntailment} & Entailment: A dog engages in a water activity and exhibits behaviors such as panting and shaking. \\ \midrule
\textbf{ClothoMoment} & A dog is barking at the heavy downpour of rain. [13s, 38s] \\ \midrule
\textbf{ClothoV2GPT} & A door creaks while a dog barks in the distance. \\ \midrule
\end{tabular}
}
\end{table}

\begin{table}[ht]
\scalebox{0.88}{
\begin{tabular}{@{}p{2.5cm}p{6.7cm}@{}}
\toprule
\textbf{CompAR} & Based on the audio elements provided, infer the ambiance or mood conveyed by this outdoor setting. Be sure to consider the impact of the specific animal and human sounds. The combination of bird chirping, insect sounds, and a human voice suggests a serene, possibly  pastoral setting. The barking dog adds to the perception of a lively yet relaxed ambiance. \\ \midrule
\textbf{DAQA} & Was the sound of the car passing by louder than the sound of the dog making noise? yes \\ \midrule
\textbf{ESC50} & dog \\ \midrule
\textbf{EpidemicSoundEffects} & the sounds of ambience, subdivision, night, insects, distant, dog, loop, bugs, and distant dog. \\ \midrule
\textbf{EzAudioCaps} & A woman speaks while a dog barks in a domestic setting, creating a lively atmosphere. \\ \midrule
\textbf{FAVDBench} & A man commands a dog. Occasionally, you may hear a dog barking in the distance \\ \midrule
\textbf{FSD50k} & 20061012. a distant ambulance siren along with some dog barks. The noise of the siren ends abruptly, presumably upon arrival to  destination. ME62+AKGCK94 into iRiver H120 decoded to midside stereo / una abulancia lejana con algun ladrido de perro \\ \midrule
\textbf{FindSounds} & dog \\ \midrule
\textbf{FreeToUseSounds} & animals dog bark med 004 \\ \midrule
\textbf{Freesound} & Creek dogs. \\ \midrule
\textbf{LAION630k} & Village quiet insects birds dogs and light activity away 140301 04 \\ \midrule
\textbf{LASS} & a lot of insects are chirping and a dog is barking. \\ \midrule
\textbf{MACS} & adults speaking while a dog barking \\ \midrule
\textbf{MULTIS} & What is happening to the dog? The dog sounds like it is upset, possibly whimpering because of hunger, pain, or fear. \\ \midrule
\textbf{NonSpeech7k} & - \\ \midrule
\textbf{Paramount} & - \\ \midrule
\textbf{PicoAudio} & dog barking at 3.038-5.038, 5.585-7.585 \\ \midrule
\textbf{ProSoundEffects} & Welsh Hillside With Occasional Creaking Trees, Birds \& Distant Sheepdog \\ \midrule
\textbf{RDATS} & A loud wheel rolling precedes a dog barking, amidst the background of crickets chirping. \\ \midrule
\textbf{SonnissGameEffects} & Village quiet insects birds dogs and light activity away 140301 04 \\ \midrule
\textbf{SonycUST} & Manhattan, medium engine, large engine, impact machinery, non-machinery impact, people talking, large crowd, dog barking/whining \\ \midrule
\textbf{SoundBible} & A dog is barking. \\ \midrule
\textbf{SoundDescs} & Common (Northern) Flicker (Colaptes Auratus) - medium close-up calls. Calls from Killdeer flying over lake, dogs barking in part  and insects in background. \\ \midrule
\textbf{SoundJay} & - \\ \midrule
\textbf{SoundVECaps} & A festive atmosphere fills the indoor setting as a person holds a small dog, and then music plays in the background while a man  speaks. \\ \midrule
\textbf{SoundingEarth} & Dandeli jungle camp, Pradhani, Karnataka 581363, Inde - Dogs howling. India, Karnataka, Dandeli, Pradhani, jungle, camp, night, dogs, insects, Binaural, Fostex FR-2LE, MS-TFB-2 microphones \\ \midrule
\textbf{SpatialSoundQA} & Identify the sound events in the audio clip. animal; dog; domestic animals, pets; grunt; pig; speech \\ \midrule
\textbf{Syncaps} & Helicopter flying before dog barking \\ \midrule
\textbf{TextToAudioGrounding} & a dog is growling and barking \\ \midrule
\textbf{VGGSound} & dog growling \\ \midrule
\textbf{WavCaps} & People are speaking and dogs are barking, with ticking sounds in the background. \\ \midrule
\textbf{WavText5K} & barking dogs. \\ \midrule
\textbf{WeSoundEffects} & - \\ \midrule
\textbf{Zapsplat} & Cartoon dog grunt 1 \\ \bottomrule
\end{tabular}
}
\end{table}

\end{document}